\documentclass[preprintnumbers,aps,prb,preprint,showpacs,showkeys]{revtex4}

\usepackage{graphicx}
\usepackage{amsmath}
\usepackage{amssymb}
\usepackage{amsfonts}

\begin{document}

\preprint{APS/123-QED}

\title{A nc-AFM simulator with Phase Locked Loop-controlled frequency detection and excitation}

\author{Laurent Nony\footnote{To whom correspondence should be addressed; E-mail:
laurent.nony@l2mp.fr.}} \affiliation{L2MP, UMR CNRS 6137,
Universit\'{e} Paul C\'{e}zanne Aix-Marseille III, Case 151, 13397
Marseille Cedex 20, France}

\author{Alexis Baratoff}
\affiliation{NCCR ``Nanoscale Science", University of Basel,
Klingelbergstr.~82, CH-4056 Basel, Switzerland}

\author{Dominique Sch\"{a}r}
\author{Oliver Pfeiffer}
\author{Adrian Wetzel}
\author{Ernst Meyer}
\affiliation{Institute of Physics, Klingelbergstr.~82, CH-4056
Basel, Switzerland\\
\\
Published in PHYSICAL REVIEW B \textbf{74}, 235439 (2006)}



\begin{abstract}
A simulation of an atomic force microscope operating in the
constant amplitude dynamic mode is described. The implementation
mimics the electronics of a real setup which includes a digital
Phase Locked Loop (PLL). The PLL is not only used as a very
sensitive frequency detector, but also to generate the
time-dependent phase-shifted signal which drives the cantilever.
The optimum adjustments of individual functional blocks and their
joint performance in typical experiments are determined in
details. Prior to testing the complete setup, the performances of
the numerical PLL and of the amplitude controller were ascertained
to be satisfactory compared to those of the real components.
Attention is also focussed on the issue of apparent dissipation,
that is of spurious variations in the driving amplitude caused by
the non-linear interaction occurring between the tip and the
surface and by the finite response times of the various
controllers. To do so, an estimate of the minimum dissipated
energy which is detectable by the instrument upon operating
conditions is given. This allows to discuss the relevance of
apparent dissipation which can be conditionally generated with the
simulator in comparison to values reported experimentally. The
analysis emphasizes that apparent dissipation can contribute to
the measured dissipation up to $15\%$ of the intrinsic dissipated
energy of the cantilever, but can be made negligible when properly
adjusting the controllers, the PLL gains and the scan speed. It is
inferred that the experimental values of dissipation reported
cannot only originate in apparent dissipation, which favors the
hypothesis of ``physical" channels of dissipation.
\end{abstract}

\maketitle

\section{Introduction}
Since almost a decade, non-contact atomic force microscopy
(nc-AFM) has proven capable of yielding images showing contrasts
down to atomic scale on metals, semiconductors, as well as
insulating ionic crystals, with or without metallic or adsorbate
overlayers~\cite{morita02a,garcia02a,giessibl03a}.  Like other
scanning force methods, the technique relies on a micro-fabricated
tip grown at the end of a cantilever. However, unlike the widely
used contact or the tapping modes, the cantilever deflection is
neither static nor driven at constant frequency, but is driven at
a frequency $\widetilde{f_0}=\widetilde{\omega_0}/2\pi$ equal to
its fundamental bending resonance frequency, slightly shifted by
the tip-sample interaction. A sufficiently large oscillation
amplitude prevents snap into contact.  A quality factor exceeding
$10^4$, readily achieved in UHV, together with frequency detection
by demodulation provide unprecedented force
sensitivity~\cite{albrecht91a, duerig92a}. A phase-locked loop
(PLL) is typically used for that purpose. Since $\widetilde{f_0}$
varies with the tip-surface distance, it deviates from $f_0$, the
fundamental bending eigenfrequency of the free cantilever.  Upon
approaching the surface, the tip is first attracted, in particular
by Van der Waals forces, which decrease $\widetilde{f_0}$. The
negative frequency shift, $\Delta f=\widetilde{f_0}-f_0$, varies
rapidly with the minimum tip-distance $d$, usually as $d^{-n}$
with $n\geq1.5$, and then as $\exp(-d/\lambda)$ a few
angstr\"{o}ms above the surface, owing to short-range chemical
and/or steric forces~\cite{guggisberg00a}. When $\Delta f$ is used
for distance control, contrasts down to the atomic scale can be
achieved. Another specific feature of the nc-AFM technique is that
the oscillation amplitude $A$ is kept constant while approaching
or scanning the surface at constant $\Delta f$.

Controlling the phase of the excitation so as to maintain it on
resonance and to make the frequency matching a preset
$\widetilde{f_0}$-value, as well as the driving amplitude so as to
keep the tip oscillation amplitude constant, respectively,
requires dedicated electronic components. Amplitude control is
usually achieved using a proportional integral controller (PIC),
hereafter referred to as APIC, whereas phase and frequency control
can be performed in two ways. In both cases the AC deflection
signal of the cantilever is filtered, then phase-shifted and
multiplied by the APIC output and by a suitable gain. The most
common method consists in using a band-pass filtered deflection
signal~\cite{gauthier02a, couturier03a}. This is referred to as
the self-excitation mode. The second method, extensively analyzed
hereafter, consists in using the PLL to generate the
time-dependent phase of the excitation signal. The PLL output is
driven by the AC deflection signal and phase-locked to it,
provided that the PLL settings are properly adjusted. Then, the
PLL continuously tracks the oscillator frequency $\widetilde{f_0}$
with high precision. Moreover, the phase lag introduced by the PLL
itself can be compensated. For reasons of clarity, this mode will
be referred to as the PLL-excitation mode. The choice of the PLL
as the excitation source has initially been motivated to take
benefit of the noise reduction due to PLLs~\cite{BEST}. A further
advantage is that the noise reduction does not only optimize the
detection of the frequency shift, but also the excitation signal.
In both modes, the phase shifter is adjusted so as the phase lag
$\varphi$ between the excitation and the tip oscillation equals
$-\pi/2$~rad throughout an experiment. If all adjustments and
controls were perfect, the oscillator would then always remain on
resonance.

The nc-AFM technique therefore requires the simultaneous operation
of three controllers~: PLL, APIC and distance controller, which
keeps constant a given $\Delta f$ while scanning the surface.
Since the tip-surface interaction makes the dynamic of the
oscillator non-linear, the combined action of those three
controllers becomes complex. They can conditionally
interplay~\cite{gauthier02a,couturier03a} and therefore influence
the dynamics of the system. Consider for instance the time the PLL
spends to track $\widetilde{f_0}$ is long compared to the time
constant of the APIC. Then, the cantilever is no longer maintained
at $\widetilde{f_0}$, but at a frequency slightly higher or lower.
Consequently, the oscillation amplitude
drops~\cite{note_virtual11} and the APIC increases the excitation
to correct the amplitude reduction. Such an apparent loss of
energy, which can as well be interpreted as a damping increase of
the cantilever, does not result of a dissipative process occurring
between the tip and the surface, but is the consequence of the bad
tracking of $\widetilde{f_0}$. So-called apparent dissipation (or
apparent damping) remains under discussions in the nc-AFM
community, which hinders the quantitative interpretation of the
experimental proofs of dissipative phenomena on the atomic scale
over a wide variety of
samples~\cite{gotsmann99a,loppacher00a,loppacher00b,bennewitz00a,gotsmann01a,hoffmann01b,giessibl02b,hoffmann03a,pfeiffer04a,hembacher05a}.
Thus, addressing the problem of apparent dissipation turns out to
be mandatory but requires to understand the complex interplay
between controllers as well as to analyze the system time
constants. Although several models of physical dissipation,
connected or not to the conservative tip-surface interaction have
been
proposed~\cite{gauthier99a,sasaki00a,gauthier00a,duerig00b,kantorovich01b,boisgard02a,boisgard02b,trevethan04a,trevethan04b,kantorovich04a}
and reviewed~\cite{note_virtual08}, the question of apparent
dissipation in the self-excitation scheme has been addressed by
two groups~\cite{gauthier02a,couturier02a,couturier03a}.
M.~Gauthier \emph{et al.}~[\onlinecite{gauthier02a}] emphasize the
interplay between the controllers and the conservative tip-sample
interaction which, although weak, can significantly affect the
damping. They put in evidence resonance effects which can
conditionally occur in damping images upon scan speed and APIC
gains. G.~Couturier et al.~\cite{couturier03a} address a similar
problem numerically and analytically. They show that the
self-excited oscillator can be conditionally stable within a
narrow domain of $K_p$ and $K_i$ gains of the APIC, but that
consequent damping variations can as well be generated upon
conservative force steps which change the borders of the stability
domain. The results mentioned above are valid for the
self-excitation mode, but the question of apparent dissipation
remains open regarding the PLL-excitation mode. However recently,
J.~Polesel-Maris and S.~Gauthier~[\onlinecite{polesel05}] have
proposed a virtual dynamic AFM based on the PLL-excitation scheme.
Their work is targeted at images calculations including realistic
force fields obtained from molecular dynamics
calculations~\cite{note_virtual12}. Their conclusions stress the
contribution of the scanning speed and of the experimental noise
to images distortion but do not address the potential contribution
of the PLL upon operating conditions.

The goal of the present work is two-fold~: 1- providing a detailed
description of the PLL-excitation based electronics of a
home-built AFM used in our laboratory;  2- assessing the
contribution of the various controllers to the dissipation signal
and in particular the contribution of the PLL. The paper is
organized as follows. In section \ref{SECTION_OVERVIEW}, an
overview of the chart of the microscope and of the attached
electronics (\emph{cf}. fig.\ref{FIG_MICROSCOPE}), is given in
terms of blocks, namely~; oscillator and optical detection (block
1), RMS-to-DC converter (block 2), amplitude controller (block 3),
PLL (block 4), phase shifter (block 5) and tip-surface distance
controller (block 6). In section \ref{SECTION_NUMERICALSCHEME},
the detailed description of the numerical scheme used to perform
the calculations is given on the base of coupled
integro-differential equations ruling each block. Section
\ref{SECTION_MINDISSDETECT} provides an estimate of the minimum
detectable dissipation by the instrument with the goal to assess
the relevance, compared to experimental results, of the apparent
dissipation which can be conditionally generated numerically.
Section \ref{SECTION_RESULTS} reports the results. In the first
part, the simulation is validated by comparing a numerical $\Delta
f$ \emph{vs.} distance curve to the analytic expression of the
$\Delta f$ due to Morse and Van der Waals interactions which does
not take into account the finite response of the various
controllers. Then the dynamic properties of the numerical PLL and
APIC upon gains are compared to those of the real components.
Section \ref{SUBSECTION_APPARENTDAMPING} gives some examples on
how apparent dissipation can be produced upon working conditions
of the PLL. Section \ref{SUBSECTION_SCANS} finally shows scan
lines computed while varying PLL gains, scan speed and APIC gains.
A discussion and a conclusion end the article.

\section{Overview}\label{SECTION_OVERVIEW}
\subsection{Description}
\indent The electronics consists of analog and digital ($12$ bits)
circuits which are described by six interconnected main blocks
operating at various sampling frequencies ($f_s$). The highest
sampling frequency among the digital blocks is the PLL one,
$f_{s_1}=20$~MHz. The PLL electronics has initially been
developped by Ch.Loppacher~[\onlinecite{loppacher98a}].

Block $1$ represents the detected oscillating tip motion coupled
to the sample surface. In the simulation, the block is described
by an equivalent analog circuit. More generally, all the analog
parts of the electronics are described in the simulation using a
larger sampling frequency compared to $f_{s_1}$, namely
$f_{s_2}=400$~MHz. This is motivated by the ultra-high vacuum
environment within which the microscope is placed, thus resulting
in a high quality factor of the cantilever, typically $Q=30000$ at
room temperature. Besides, nc-AFM cantilevers have typical
fundamental eigenfrequencies $f_0\simeq150$~kHz. The chosen
sampling frequency should therefore insure a proper integration of
the differential equations with an error weak enough. The signal
of the oscillating cantilever motion goes into a band pass filter
which cuts-off its low and high frequencies components. The
bandwidth of the filter is typically $60$~kHz, centered on the
resonance frequency of the cantilever. Despite the filter has been
implemented in the simulation, no noise has been considered, so
far. The signal is then sent to other blocks depicting the
interconnected parts of two boards, namely an analog/digital one,
the ``PLL board", and a fully digital one which integrates a
Digital Signal Processor (DSP), the ``DSP board". The boards share
data \emph{via} a ``communication bus" operating at
$f_{s_3}=10$~kHz, the lowest frequency of the digital electronics.

Block $2$ stands for the lone analog part of the PLL board
($f_s=f_{s_2}$). It consists of a RMS-to-DC converter. The block
output is the rms value of the oscillations amplitude,
$A_\text{rms}(t)$. $A_\text{rms}(t)$ is provided to block 3, one
of the two PICs implemented on the DSP ($f_s=f_{s_3}$). When
operating in the nc-AFM mode, the block output is the DC value of
the driving amplitude which maintains constant the reference value
of the oscillations amplitude, $A_0^\text{set}$. This is why it is
referred to as the amplitude controller, APIC. For technical
reasons due to the chips, the signal is saturated between $0$ and
$10$~V.

\begin{figure}[h]
  \includegraphics[width=17cm]{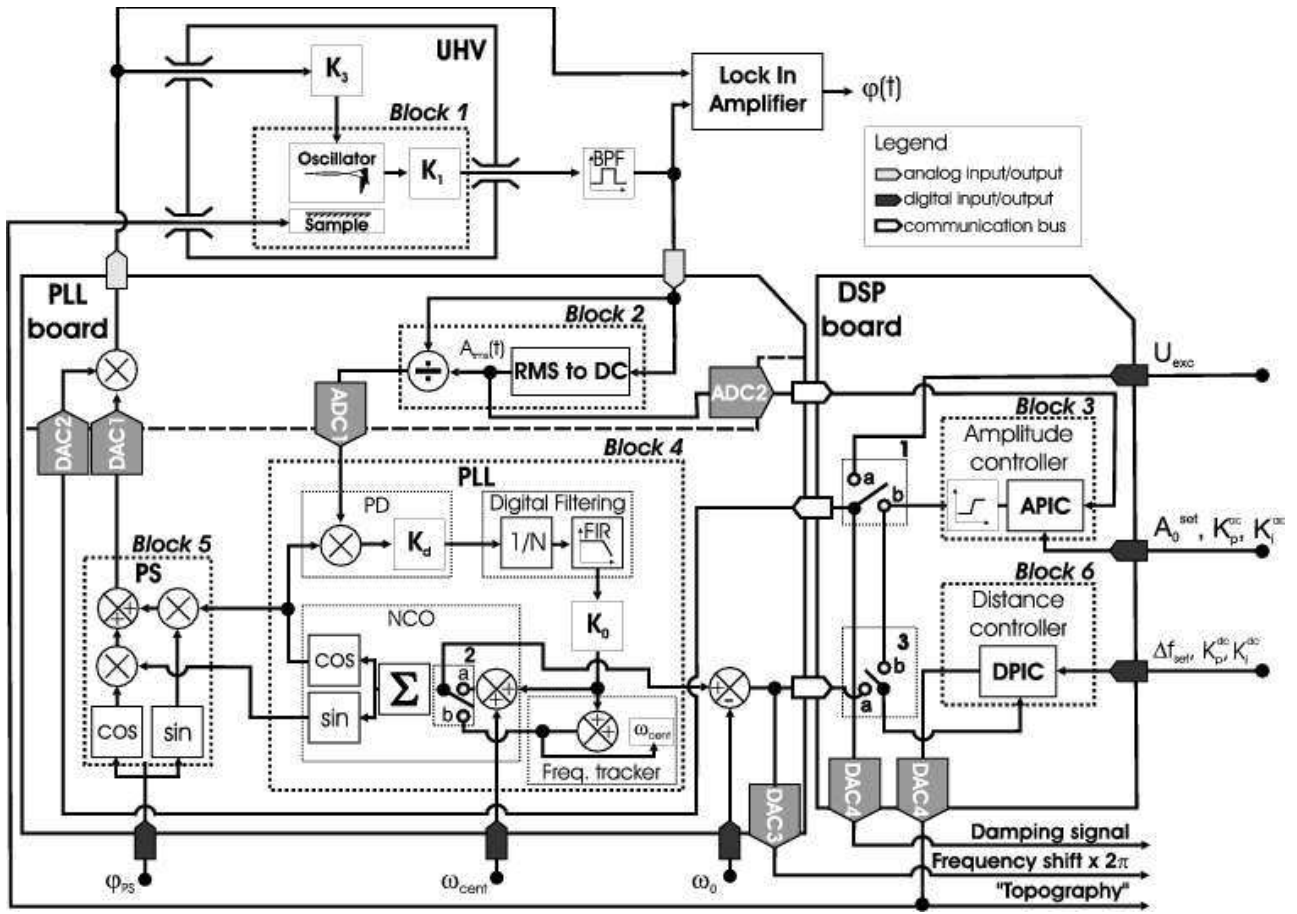}\\
  \caption{Scheme of the simulator operating in nc-AFM,
   based on the design of the electronics of the real apparatus.}\label{FIG_MICROSCOPE}
\end{figure}

The dashed line in fig.\ref{FIG_MICROSCOPE} depicts the border
between analog and digital circuits in the PLL board. The digital
PLL, block $4$ ($f_s=f_{s_1}$), consists of three sub-blocks~: a
Phase Detector (PD), a Numerical Controlled Oscillator (NCO) and a
filtering stage consisting of a decimation filter and a Finite
Impulse Response (FIR) low pass filter in series. The PLL receives
the signal of the oscillation divided by $A_\text{rms}(t)$ plus an
external parameter~: the ``center frequency",
$f_\text{cent}=\omega_\text{cent}/2\pi$. $f_\text{cent}$ specifies
the frequency to which the input signal has to be compared to for
the demodulation frequency stage. This point is particularly
addressed in section \ref{SUBSECTION_PLL}. The NCO generates the
digital $\sin$ and $\cos$ waveforms of the time-dependent phase,
$\varphi_\text{nco}(t)+\varphi_\text{pll}$(t), ideally identical
to the one of the input signal. $\varphi_\text{pll}(t)$ is
correlated to the error which is potentially produced while the
frequency demodulation, upon operating conditions. The $\sin$ and
$\cos$ waveforms are then sent to a digital phase shifter, block
$5$ ($f_s=f_{s_1}$) which shifts the incoming phase
$\varphi_\text{nco}(t)+\varphi_\text{pll}(t)$ by a constant
amount, $\varphi_\text{ps}$, set by the user. Since the cantilever
is usually driven at $\widetilde{f_0}$, $\varphi_\text{ps}$ is
adjusted to make that condition fulfilled~\cite{note_virtual13},
namely~:

\begin{equation}
\varphi_\text{nco}(t)+\varphi_\text{pll}(t)+\varphi_\text{ps}=\widetilde{\omega_0}t,
\end{equation}

\noindent Indeed, the PLL produces the phase locked to the input,
that is
$\varphi_\text{nco}(t)+\varphi_\text{pll}(t)\simeq\widetilde{\omega_0}t-\pi/2$.
If it optimally operates, $\varphi_\text{pll}(t)\simeq 0$.
$\varphi_\text{ps}$ has therefore to be set equal to $+\pi/2$ to
maintain the excitation at the resonance frequency prior to
starting the experiments. Consequently, $\varphi = -\pi/2$~rad.
The block output,
$\sin\left[\varphi_\text{nco}(t)+\varphi_\text{pll}(t)+\varphi_\text{ps}\right]$,
 is converted into an analog signal and then multiplied by the APIC output, thus generating the full AC
excitation applied to the piezoelectric actuator to drive the cantilever.

Block $6$ is the second PIC of the DSP ($f_s=f_{s_3}$). It
controls the tip-surface distance to maintain constant either a
given value of the frequency shift, or a given value of the
driving amplitude while performing a scan line (switch $3$ set to
location ``a" or ``b", respectively in fig.\ref{FIG_MICROSCOPE}).
The output is the so-called ``topography" signal. The block is
referred to as the distance controller, DPIC.

Finally, a digital lock-in amplifier detects the phase lag, $\varphi$, between the excitation signal provided to the oscillator and the
oscillating cantilever motion.

\subsection{Time considerations}
Analog and digital data are properly transformed by
Analog-to-Digital and Digital-to-Analog Converters (ADC and DAC,
respectively). In the electronics, ADC1 is an \textit{AD$9042$}
(\emph{cf.} fig.\ref{FIG_MICROSCOPE}) with a nominal sampling rate
of $41\times 10^6$ samples \emph{per} second~\cite{NOTE07}. This
ensures the analog signal is sampled quick enough and properly
operated by the PLL at $f_{s_1}$. This ADC is therefore not
described in the simulation. ADC$2$ (\textit{ADS $7805$}) has a
nominal frequency of $100$~kHz~[\onlinecite{NOTE07}]. The signal
is transmitted to the communication bus, the bandwidth of which is
ten times smaller. Its role is therefore as well supposed to be
negligible. The code is implemented assuming that the RMS-to-DC
output signal is provided to the communication bus operating at
$f_{s_3}$. DAC1 (\textit{AD $668$}) is a $12$ bits ultrahigh speed
converter. It receives the digital waveform coming from the PS.
Indeed, it must be fast enough to provide a proper analog signal
to hold the excitation. Its nominal reference bandwidth is
$15$~MHz~[\onlinecite{NOTE07}]. To make the code implementation
easier, the DAC has not been implemented neither. Thus, it is
assumed that the PS signal directly provides the signal at
$f_{s_1}$ to perform the analog multiplication, itself processed
at $f_{s_3}$ due to the APIC output. The others DACs have all
nominal bandwidths much larger than the communication bus one and
are also assumed to play negligible roles.

\section{Numerical scheme}\label{SECTION_NUMERICALSCHEME}
\subsection{Block 1: oscillator and optical detection}\label{SUBSECTION_OSCILLATOR}
The block mimics the photodiodes acquiring the signal of the
motion of the oscillating cantilever. The equation describing its
behavior is given by the differential equation of the harmonic
oscillator~:

\begin{equation}\label{EQU_OTCS}
\ddot{z}(t)+\frac{\omega_0}{Q}\dot{z}(t)+\omega_0^{2}z(t)=\omega_0^{2}\Xi_\text{exc}(t)+
\frac{\omega_0^2F_\text{int}(t)}{k_c}
\end{equation}

\noindent$\omega_0=2\pi f_0$, $Q$, $k_c$ stand for the angular
resonance frequency, quality factor and cantilever stiffness of
the free oscillator, respectively. $z(t)$, $\Xi_\text{exc}(t)$ and
$F_\text{int}(t)$ are the instantaneous location of the tip,
excitation signal driving the cantilever and the interaction force
acting between the tip and the surface, respectively. The equation
is solved with a modified Verlet algorithm, so-called leapfrog
algorithm~\cite{Rapaport95a}, using a time step $\Delta
t_{s_2}=1/f_{s_2}=5$~ns. In the followings, the time will be
denoted by its discrete notation~: $t\rightarrow t_i=i\times
\Delta t_{s_2}$.
\newline The instantaneous value of the driving amplitude $\Xi_\text{exc}(t_i)$ (units~:~m) can be written as~:

\begin{equation}\label{EQU_XIEXC(T)}
\Xi_\text{exc}(t_i)=K_3 A_\text{exc}(t_i) z_\text{ps}(t_i)
\end{equation}

\noindent $K_3$ (units~:~m.V$^{-1}$) represents the linear
transfer function of the piezoelectric actuator driving the
cantilever. $A_\text{exc}(t_i)$ (units~:~V) is the APIC output
(\emph{cf.} section \ref{SUBSECTION_APIC}). It is proportional to
the damping signal according to~:

\begin{equation}\label{EQU_AEXC(T)}
K_3A_\text{exc}(t_i)=\frac{\Gamma(t_i) A_0}{\omega_0},
\end{equation}

\noindent $\Gamma(t_i)$ and $A_0$ (units~:~s$^{-1}$ and m,
respectively) being the damping signal and oscillations amplitude
of the cantilever when driven at $f_0$, respectively. When the
cantilever is externally driven and if no interaction occurs,
$A_\text{exc}(t_i)$ can be written as a function of $A_0$ and of
the quality factor of the cantilever~:

\begin{equation}\label{EQU_AEXCFREE}
K_3A_\text{exc,0}=\frac{A_0}{Q}
\end{equation}

\noindent Then the damping of the free cantilever equals~:

\begin{equation}\label{EQU_DAMPFREE}
\Gamma_0=\frac{\omega_0}{Q}
\end{equation}

\noindent In nc-AFM, the dissipation is commonly expressed in terms of dissipated energy \emph{per} oscillation cycle, $E_{d_0}$. For a cantilever
with a high quality factor oscillating with an amplitude $A_0$~:

\begin{equation}\label{EQU_ENERGYLOSSINTRINSIC}
    E_{d_0}(A_0)=\frac{\pi k_c A_0^2}{Q}=\frac{\pi k_c A^2\Gamma_0}{\omega_0}
\end{equation}

\noindent In UHV and at room temperature, $Q=30000$. Besides,
nc-AFM commercial cantilevers have typical
stiffnesses~\cite{note_virtual03} $k_c \approx 40$~N.m$^{-1}$.
Considering $A_0=10$~nm, the intrinsic dissipated energy
\emph{per} cycle of the cantilever is then $E_{d_0}\simeq
2.6$~eV/cycle.

In equation \ref{EQU_XIEXC(T)}, $z_\text{ps}(t_i)$ is the AC part of the excitation signal (\emph{cf.} section \ref{SUBSECTION_PS}). It is
provided by the PS when the PLL is engaged. When the steady state is reached, \emph{e.g.} $t_i\gg t_\text{steady}\simeq 2Q/f_0$, the block output
is~:

\begin{equation}\label{EQU_Z(T)}
K_1 z(t_i)=K_1 A(t_i)\sin\left[\omega t_i+\varphi(t_i)\right]
\end{equation}

\noindent $K_1$ (V.m$^{-1}$) depicts the transfer function of the photodiodes which is assumed to be linear within the bandwidth ($3$~MHz in the
real setup). If the damping is kept constant, the amplitude and the phase, $A(t_i)$ and $\varphi(t_i)$ respectively, are supposed to be constant
as well. This is no longer true once the various controllers are engaged, therefore their time dependence is explicitly preserved.

In equation \ref{EQU_OTCS}, the interaction force
$F_\text{int}(r)=-\partial_{r}V_\text{int}(r)$ is derived from a
conservative potential consisting of two components~: a long-range
part, depicted by a Van der Waals term defined between a sphere
and a half-plane and a short-range part, prevailing at closer
distances, depicted by a Morse potential~:

\begin{equation}\label{EQU_FORCE}
 V_\text{int}(r)=-\frac{HR}{6r}-U_0 \left[ 2
e^{-\frac{r-r_c}{\lambda}}
-e^{-\frac{2\left(r-r_c\right)}{\lambda}}\right]
\end{equation}

\noindent $H$ and $R$ are the Hamaker constant of the
tip-vacuum-surface interface and tip's radius, respectively.
$U_0$, $r_c$ and $\lambda$ are the depth, equilibrium position and
range of the Morse potential. The instantaneous tip-surface
separation is $r(t_i)=D(t_i)-z(t_i)$, where $D(t_i)$ is the
distance between the surface location and the cantilever position
at rest. So far, neither elastic deformation of the sample and
tip, nor dissipative interaction have been considered.

The signal $K_1 z(t_i)$ then gets into the band pass filter (BPF), the central frequency of which, $f_c=\omega_c/2\pi$, equals the resonance
frequency of the cantilever, $f_0$, with a bandwidth $B_W\simeq 60$~kHz. The output, $z_\text{bpf}(t_i)$ (units~:~V), is ruled by~:

\begin{equation}\label{EQU_ZBPF(T)}
\ddot{z}_\text{bpf}(t)+2\pi
B_W\dot{z}_\text{bpf}(t)+\omega_c^{2}z_\text{bpf}(t)=2\pi B_W
 K_1 \dot{z}(t)
\end{equation}

\noindent $z_\text{bpf}(t_i)$ is then provided to the RMS-to-DC
converter of the PLL board.

\subsection{Block 2: RMS-to-DC converter}\label{SUBSECTION_RMS2DC}
The converter is the only analog part of the PLL board. The
related differential equation is integrated at $f_{s_2}$. The chip
(AD734) computes the square root of the squared value of the
incoming signal, preliminary filtered by a first-order low pass
filter, the cut-off frequency of which is $f_{co}=400$~Hz. The
output is the amplitude (DC value) of the oscillation,
$A_\text{rms}(t_i)$ (units~:~V)~:

\begin{equation}\label{EQU_RMS2DC}
A_\text{rms}(t_i)=\sqrt{V_s(t_i)},
\end{equation}

\noindent $V_s(t_i)$ being the output of the first-order low pass
filter~:

\begin{equation}\label{EQU_RMS2DCLP}
\tau_\text{rms}\dot{V}_s(t)+V_s(t)=z_\text{bpf}^2(t_i),
\end{equation}

\noindent with $\tau_\text{rms}=1/(2\pi f_{co})\simeq 400$~$\mu$s.

$z_\text{bpf}(t_i)$ is then divided by $A_\text{rms}(t_i)$ in order to normalize the amplitude of the waveform. The signal thus normalized is sent
to the ADC1 to be operated by the digital PLL.

\subsection{Block 3: amplitude controller}\label{SUBSECTION_APIC}
The block represents a digital PI controller implemented in the DSP board. The controller receives the RMS-to-DC output signal \emph{via} the
communication bus. Since the bus operates at $f_{s_3}=10$~kHz, the time step used to solve the related differential equation is $\Delta
t_{s_3}~=~1/f_{s_3}~=~100$~$\mu$s. Besides $A_\text{rms}$, the controller receives three external parameters~: the proportional and integral
gains, $K_p^\text{ac}$ and $K_i^\text{ac}$ respectively (units~:~dimensionless and s$^{-1}$, respectively), and the reference amplitude expected
to be kept constant as soon as the controller is engaged (switch $1$ set to location ``b" in fig.\ref{FIG_MICROSCOPE}), $A_0^\text{set}$
(units~:~V). The block output is the DC value of the excitation, previously referred to as $A_\text{exc}(t_i)$ (\emph{cf.}
equ.\ref{EQU_XIEXC(T)})~:

\begin{align}\label{EQU_APIC}
A_\text{exc}(t_{i})= &
K_p^\text{ac}\left[A_0^\text{set}-A_\text{rms}(t_i)\right] \notag \\
& +
\sum_{k=0}^{i}{K_i^\text{ac}\left[A_0^\text{set}-A_\text{rms}(t_k)\right]\Delta
t_{s_3}}
\end{align}

\noindent Engaging the APIC makes the nc-AFM mode effective. This requires the PLL-excitation mode (block 4, \emph{cf.} section
\ref{SUBSECTION_PLL}) to be already engaged. If operating at $\widetilde{f_0}$, then $A_\text{rms}/K_1$ equals the resonance amplitude, $A_0$.
$A_\text{exc}$ is then minimal.

\subsection{Block 4: PLL}\label{SUBSECTION_PLL}
Before starting this section, note that some of the elements
detailed hereafter are adapted from the book by
R.Best~[\onlinecite{BEST}]. The digital PLL consists of three
sub-units~: a Phase Detector, a decimation filter and a FIR low
pass filter in series and a NCO. The block operates at
$f_{s_1}~=~20$~MHz, with the related time step $\Delta
t_{s_1}=1/f_{s_1}=50$~ns. In the electronics, various FIR low pass
filters have been implemented upon the desired sensitivity in the
frequency detection, among which a $19^\text{th}$ order filter
with a 3~kHz cut-off frequency and a $45^\text{th}$ order filter
with a 500~Hz cut-off frequency. Both of them can be used in the
simulation.

\subsubsection{Phase detector}\label{SUBSUBSECTION_PD}
The PD is analogous to a multiplier regarding the two input
signals~: the BPF output divided by $A_\text{rms}$ and the $\cos$
waveform coming out of the NCO (\emph{cf.}
fig.\ref{FIG_MICROSCOPE}). Their product is multiplied by a
further gain, $K_d$ (units~: V) converting the dimensionless
signal into volts to be operated by the FIR low pass filter. The
instantaneous block output is referred to as $K_d z_e(t_i)$~:

\begin{equation}\label{EQU_PD}
K_d z_e(t_i)=K_d \frac{z_\text{bpf}(t_i)}{A_\text{rms}(t_i)}
\cos\left(\varphi_\text{nco}(t_i) \right)
\end{equation}

\subsubsection{Filtering stage}\label{SUBSUBSECTION_FIRFILTER}
Assume that $\widetilde{\omega_0}(t_i)$ and
$\omega_\text{nco}(t_i)$ are the instantaneous angular frequencies
of the cantilever and of the signal generated by the NCO,
respectively. $K_d z_e(t_i)$ consists of a high frequency
component~: $\widetilde{\omega_0}(t_i)+\omega_\text{nco}(t_i)$ and
a low frequency one~: $\delta
\omega(t_i)=\widetilde{\omega_0}(t_i)-\omega_\text{nco}(t_i)$. The
FIR low pass filter cuts off the high frequency component and
produces $u_f(t_i)~\propto~\sin\left\{\delta
\omega(t_i)t_i\right\}~\propto~ \left[\delta
\omega(t_i)\right]\times t_i$, which can be referred to as an
error signal of the PLL. Indeed, when the PLL optimally operates,
$\omega_\text{nco}(t_i)$ almost perfectly matches
$\widetilde{\omega_0}(t_i)$. The instantaneous value $u_f(t_i)$
can therefore be interpreted as a correction term in the PLL
cycle.

Before being operated by the FIR low pass filter, the signal is
processed by the decimation filter. The filter averages
$K_dz_e(t_i)$ over $N_{ds}$ PLL cycles upon the FIR low pass
filter cut-off frequency. For instance, $N_{ds}=400$ for the 3~kHz
low pass filter. The updating rate of the FIR low pass filter is
therefore $f_{s_1}/N_{ds}=50$~kHz. The digital data are averaged
over those $N_{ds}$ cycles. The average value is fed at the first
entry of a buffer $B$ consisting of $N_\text{fir}$ entries. The
entries of the buffer are then all shifted by one into the buffer.
At a given moment in time, $t_i$, $u_f(t_i)$ is given by the
following algorithm~:

\begin{equation}\label{EQU_FILTERINGSTAGEPLL}
\left\{
\begin{array}{ll}
& B_k=\frac{\sum_{j=k-N_{ds}}^{N_{ds}} K_d z_e(t_j)}{N_{ds}} \\
& \text{shift of the buffer entries}\\
& u_f(t_i)=\sum_{k=i-N_\text{fir}}^i c_k \times B(t_k)
\end{array}
\right.
\end{equation}

\noindent $N_\text{fir}$ is the order of the FIR low pass filter
($N_\text{fir}\ll N_{ds}$) and $c_k$ is the k$^{th}$ coefficient
of the FIR low pass filter. Once the buffer is transmitted, it is
initialized and filled again. Finally, $u_f(t_i)$ is multiplied by
a further gain, $K_0$, which depicts the linear conversion of the
signal from volts to rad.s$^{-1}$ (units: rad.V$^{-1}$.s$^{-1}$)
and provided to the NCO.

\subsubsection{Numerical Controlled Oscillator}\label{SUBSUBSECTION_NCO}
We first assume that the frequency tracker of the PLL is disengaged (switch $2$ set to location ``a" in fig.\ref{FIG_MICROSCOPE}). Its role is
carefully addressed in section \ref{SUBSUBSECTION_FTRACKER}. The NCO adds the instantaneous angular frequency $K_0 u_f(t_i)$ to an external input,
the center angular frequency of the PLL, $\omega_\text{cent}$. $\omega_\text{cent}$ is fixed equal to the angular resonance frequency of the free
cantilever $\omega_0$, prior to starting the experiments. The signal is then integrated, which produces the related phase,
$\varphi_\text{nco}(t_i)$, locked to the one of the cantilever~:

\begin{equation}\label{EQU_FINCO}
\varphi_\text{nco}(t_i)= \sum_{k=\text{pll}}^{i}
\left[\omega_\text{cent}+K_0u_f(t_k) \right] \Delta t_{s_1},
\end{equation}

\noindent $t_\text{pll}$ being the moment when the PLL is engaged. Obviously, the PLL has to be engaged once the oscillator has reached its steady
state and before the APIC.

\subsubsection{Frequency demodulation}\label{SUBSUBSECTION_FDEMODULATION}
When the tip is located far from the surface,
$\widetilde{\omega_0}=\omega_0$. Once approached close enough from
it, $\widetilde{\omega_0}$ starts decreasing. Meanwhile, the NCO
produces $\omega_\text{nco}(t_i)=\omega_\text{cent}+K_0u_f(t_i)$,
as mentioned above. When the frequency tracker is disengaged,
$\omega_\text{cent}$ is kept constant and matches the resonance
frequency of the free cantilever, $\omega_\text{cent}=\omega_0$.
Therefore $K_0 u_f(t_i)$ is nothing but the instantaneous
frequency shift (actually $2\pi\times \Delta f$) of the tip
interacting with the surface. In other words~:

\begin{equation}\label{EQU_DELTAF}
\omega_\text{cent}+K_0 u_f(t_i)-\omega_0=2\pi \Delta f(t_i)
\end{equation}

\noindent $K_0$ is a key parameter of the PLL. It sets its
capability to get locked to the input signal and in turn it sets
its stability. R.Best defines $K_0$ from the locking range $\Delta
\omega_l$ of the PLL, \emph{e.g.} the frequency gap with respect
to the center frequency the PLL can detect remaining
locked~\cite{BEST}. On the hardware level, the control signal
$u_f(t)$ is limited to a range which is smaller than the supply
voltages, usually $\pm5$V. Assuming $u_{f_\text{m}}$ and
$u_{f_\text{M}}$ be the minimum and maximum values allowed for
$u_f$, Best defines $K_0$ as~:

\begin{equation}\label{EQU_K0BEST}
K_0=\frac{3\Delta \omega_l}{u_{f_\text{M}}-u_{f_\text{m}}}
\end{equation}

\noindent Therefore $K_0$ is related to the maximum frequency
shift detectable \emph{per} volt within the detection range of the
low pass filter. Practically, the value of $K_0$ is not accessible
\textit{a priori}. It's easier to set the locking range $\Delta
\omega_l$. For an oscillation at $f_0=150$~kHz, frequency shifts
of about a few hundreds of hertz are typically
expected\cite{note_virtual16}. We can therefore choose the 3~kHz
FIR low pass filter to insure a proper detection of $\Delta f$,
which sets the locking range to $\Delta \omega_l=2\pi \times
6000$~rad.s$^{-1}$. The maximal value of $K_0$ expected is then
$\simeq 11000$~rad.V$^{-1}$.s$^{-1}$, which is an excellent
estimate as detailed in section \ref{SECTION_RESULTS}.

\subsubsection{Frequency tracker}\label{SUBSUBSECTION_FTRACKER}
The frequency tracker is a specific feature of our digital PLL.
When engaged (switch $2$ set to location ``b" in
fig.\ref{FIG_MICROSCOPE}), the center frequency is continuously
updated by the FIR low pass filter output~:

\begin{equation}\label{EQU_OMEGACENTFREQTRACKER}
\omega_\text{cent}(t_i)=\omega_\text{cent}(t_{i-1})+K_0u_f(t_i)
\end{equation}

\noindent The updating frequency is 2.5~kHz. The frequency tracker
has been implemented in order to compensate the fact that the
frequency demodulation was performed \emph{via} the lone
proportional gain $K_0$. Thus, as mentioned before, $K_0u_f(t_i)$
can be interpreted as the error signal produced in the frequency
detection compared to $\omega_\text{cent}$. Consequently, this
error is also integrated by the NCO, which leads to an additional
phase lag added to $\varphi_\text{nco}$ at each PLL cycle and
previously referred to as $\varphi_\text{pll}$.
$\varphi_\text{pll}$ \emph{per} PLL cycle can approximately be
estimated to~:

\begin{equation}\label{EQU_PHIPLL}
\Delta
\varphi_\text{pll}=\varphi_\text{pll}(t_{i+1})-\varphi_\text{pll}(t_i)\approx
K_0u_f(t_i) \times \Delta t_{s_1}
\end{equation}

\noindent $\varphi_\text{pll}$ would be zero if no frequency shift occurred, which is the case in most of the applications using PLLs. But while
approaching, $\Delta f$ decreases continuously, therefore so does $\varphi_\text{pll}$. On the contrary, when the frequency tracker is engaged,
$\omega_\text{cent}$ is continuously updated. The error in the frequency detection drops to zero. More exactly, it is equal to the difference
between two consecutive values of $\omega_\text{cent}$~: $\epsilon \simeq \omega_\text{cent}(t_i)-\omega_\text{cent}(t_{i-1})$, but is necessarily
small and so is $\varphi_\text{pll}$.

\begin{figure}
  \includegraphics[height=8cm]{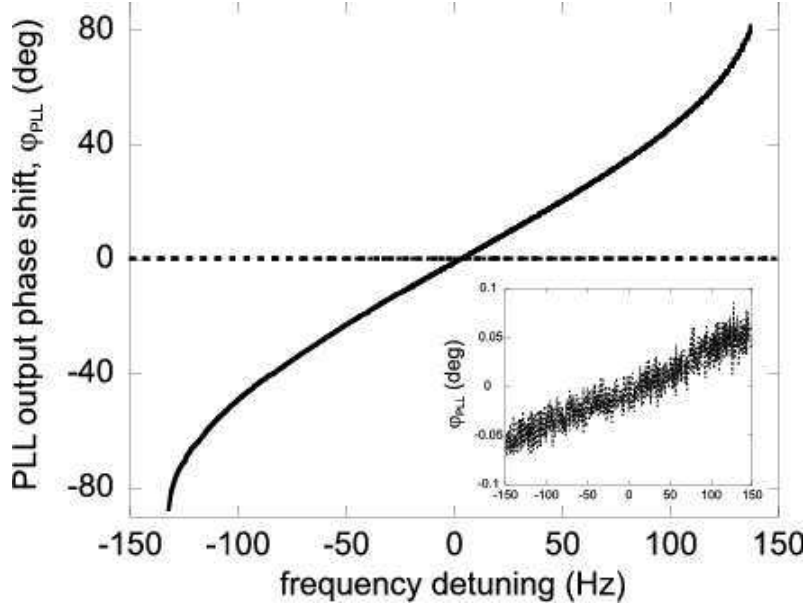}\\
  \caption{Phase shift between a $150$~kHz sinusoidal waveform sent to the real PLL and the PLL output
  waveform ($f_\text{cent}=150$~kHz) when tuning the input frequency from $-150$ to $+150$~Hz
  upon the frequency tracker is engaged or not. The 3kHz FIR low pass filter has been
  used. When the tracker is disengaged, the phase lag can reach $\pm 80$~degrees (continuous black line,
  amplitude of the input waveform~: $A_w=110$~mV peak-to-peak). When it is engaged, the phase lag drops to almost
  zero (dotted black line and inset, similar input waveform). When the input frequency accurately matches the center
  frequency, the shift is zero (modulo an error corresponding to a few fractions of degrees as shown in the inset).}\label{FIG_FREQUTRACK}
\end{figure}

To assess how sensitive to frequency changes the phase is, the
following experiment is carried out. A $150$~kHz sinusoidal
waveform is generated by means of a function generator and sent to
the real PLL. The frequency is then slowly detuned from $-150$~Hz
up to $+150$~Hz. The phase lag between input and output waveforms,
$\varphi_\text{pll}$, is recorded with a lock-in amplifier (Perkin
Elmer 7280) and reported as a function of the detuning. The PLL
center frequency is fixed to $f_\text{cent}=150$~kHz. The
experiment is repeated the frequency tracker being engaged and
disengaged. Two amplitudes of the PLL input waveform are used. In
this experiment, the input waveform stands for the oscillatory
motion of the cantilever and the tuning for the shift occurring
when the tip is approached towards the surface upon attractive or
repulsive forces. The results are reported in
fig.\ref{FIG_FREQUTRACK}. When the tracker is disengaged, the
maximum detuning corresponds to a phase lag of $\pm 80$~degrees,
which means that the cantilever would then be driven off resonance
severely. On the opposite, when engaged, the phase lag reduces
(inset) to $\pm 0.05$ degree.

This feature has no consequence when the PLL is only used as a
frequency demodulator like in the self-excitation mode. On the
opposite in the PLL-excitation scheme, this point is crucial since
the PLL is produces the excitation signal. Therefore particular
attention has to be paid on the way it is produced. If it is
abnormally phase shifted, then the oscillation amplitude drops and
consequently apparent dissipation is generated, as shown in
section \ref{SECTION_RESULTS}.

\subsection{Block 5: phase shifter}\label{SUBSECTION_PS}
The PS receives the $\sin$ and $\cos$ waveforms generated by the
NCO. A further input to the block is the phase lag,
$\varphi_\text{ps}$, fixed prior to starting the experiments to
make the cantilever oscillating at $f_0$. The PS digitally
computes~:

\begin{align}\label{EQU_ZPS(T)}
z_\text{ps}(t_i) & =
\sin\left[\varphi_\text{nco}(t_i)+\varphi_\text{pll} \right] \times \cos\left(\varphi_\text{ps} \right) \notag \\
                 & \text{~~~~~~~~~~~~~~~~~~~} +\cos\left[\varphi_\text{nco}(t_i)+\varphi_\text{pll}\right] \times \sin \left(\varphi_\text{ps} \right) \notag \\
                 & = \sin\left[\varphi_\text{nco}(t_i)+\varphi_\text{pll}+\varphi_\text{ps}\right]
\end{align}

\noindent When the system is being operated in the PLL-excitation
mode, $z_\text{ps}(t)$ is converted into an analog signal by the
DAC1 and multiplied by the APIC output.

\subsection{Block 6: distance controller}\label{SUBSECTION_DPIC}
The distance controller is the second digital PI controller implemented in the DSP operating at $f_{s_3}$. The block gets the setpoint value of
the signal ($\Delta f$ or damping) onto which the control of the tip-sample distance is performed and the proportional and integral gains,
$K_p^\text{dc}$ and $K_i^\text{dc}$, respectively. Here, let's assume that the reference signal is the frequency shift, as depicted in
fig.\ref{FIG_MICROSCOPE}. We have arbitrarily chosen not to describe the transfer function of the z-piezo drive. Therefore $K_p^\text{dc}$ and
$K_i^\text{dc}$ have natural units (nm.Hz$^{-1}$ and nm.Hz$^{-1}$.s$^{-1}$, respectively). The controller is described by~:

\begin{align}\label{EQU_DPIC}
D(t_{i}) = & D(t_\text{dc}) + K_p^\text{dc} \left[ \Delta
f_\text{set}-\Delta f(t_i)\right] \notag \\
+& \sum_{k>\text{dc}}^{i}{K_i^\text{dc}\left[\Delta
f_\text{set}-\Delta f(t_k)\right]\Delta t_{s_3}},
\end{align}

\noindent $D(t_\text{dc})$ being the tip-surface distance when the DPIC is engaged.

\subsection{Lock-in amplifier}\label{SUBSECTION_LOCKIN}
The description of the lock-in amplifier implemented in the simulation does not depict the detailed operational mode of the real lock-in which is
used to monitor the phase shift of the oscillator (Perkin Elmer 7280). Its purpose is to provide an easy way to estimate the phase shift between
the excitation and the oscillation. The calculation of the phase is performed at $2.5$~kHz. The buffer used to extract the phase therefore
consists of $n_\text{lock-in}=f_{s_1}/2.5\text{~kHz}=8000$ samples. The numerical code used to describe it is~:

\begin{align}\label{EQU_FILOCKIN}
&\tan(\varphi(t_i))= \notag\\
&\frac{\sum_{k=i-n_\text{lock-in}}^{i}z_\text{bpf}(t_k)\times
\sin[\varphi_\text{nco}(t_k)+\varphi_\text{pll}+\varphi_\text{ps}]}{\sum_{k=i-n_\text{lock-in}}^{i}z_\text{bpf}(t_k)\times
\cos[\varphi_\text{nco}(t_k)+\varphi_\text{pll}+\varphi_\text{ps}]}
\end{align}

\subsection{Code implementation}\label{SUBSECTION_CODEIMPLEMENTATION}
The numerical code has been implemented with LabView$^\text{TM}$
$6.1$, supplied by National Instruments$^\text{TM}$. It consists
of a user interface where all the parameters are tunable at
run-time, like during a real experiment. The couple of
integro-differential equations \ref{EQU_OTCS}, \ref{EQU_XIEXC(T)},
\ref{EQU_ZBPF(T)}, \ref{EQU_RMS2DC}, \ref{EQU_RMS2DCLP},
\ref{EQU_APIC}, \ref{EQU_PD}, \ref{EQU_FILTERINGSTAGEPLL},
\ref{EQU_FINCO}, \ref{EQU_ZPS(T)} and \ref{EQU_DPIC} are
integrated at their respective sampling frequencies. The monitored
signals are the oscillation amplitude $A_\text{rms}$
(equ.\ref{EQU_RMS2DC}), the frequency shift $\Delta f$
(equ.\ref{EQU_DELTAF}), the phase $\varphi$
(equ.\ref{EQU_FILOCKIN}) and the relative damping
$\Gamma/\Gamma_0-1$=$QK_3A_{exc}/A_0-1$, deduced from the APIC
output (equ.\ref{EQU_AEXC(T)}). The connection to the dissipated
energy \emph{per} cycle $E_d$ is given by equation
\ref{EQU_ENERGYLOSSINTRINSIC}, that is
$\Gamma/\Gamma_0-1=E_d/E_{d_0}-1$.

\section{Apparent dissipation \emph{vs.} minimum dissipation}\label{SECTION_MINDISSDETECT}
Addressing the question of apparent dissipation requires to
estimate the minimum dissipation which is detectable by the
instrument upon operating conditions. Beyond the specificities of
the PLL- or self-excitation modes, important parameters like
quality factor $Q$, temperature and bandwidth of the measurement
must be considered.

We here focus on the minimum dissipated energy, $\delta E_d$, due to thermal fluctuations of the cantilever when it oscillates close to a surface.
Thermal driving forces are connected to the energy dissipation by the $Q$ factor of the cantilever. The thermal kicks introduce fluctuations of
amplitude and phase and therefore fluctuations of the energy dissipation. This is true for a free cantilever, but the contribution of the thermal
noise is expected to be even more pronounced when the tip is close to the surface. Then, the fluctuations of the interaction force $\delta
F_\text{int}$ have a strong influence on the nonlinear dynamics of the cantilever, in particular when the tip is at distances involving
short-range forces where the nonlinearity is more pronounced.

The instrumental noise (\emph{cf.} Ch.2 in refs.[\onlinecite{morita02a}] and [\onlinecite{polesel05}]), essentially due to electronic components,
is not considered and we further assume that the electronic blocks (RMS-to-DC, PI controllers, PLL) operate perfectly. Doing so, $\delta E_d$ is
under-estimated but the framework of this section is to provide a ground value to be compared to the values obtained with the simulation.

\subsection{Connection between $\delta E_d$ and $\delta F_\text{int}$}

The fluctuation of the dissipated energy \emph{per} cycle can be connected to the fluctuation of damping $\delta \Gamma$, \emph{via}
equ.\ref{EQU_ENERGYLOSSINTRINSIC}~:

\begin{equation}\label{}
    \delta E_d=\pi k_c A_0^2 \frac{\delta \Gamma}{\omega_0}
\end{equation}

\noindent Besides, because $A_\text{exc}=A_0/Q=A_0\Gamma_0/\omega_0=F_\text{exc}/k_c$ on resonance and because the tip-sample interaction force
$F_\text{int}$ can be treated, to first order~\cite{note_virtual17}, on the same level as $F_\text{exc}$, a fluctuation of $F_\text{int}$ should
produce, a fluctuation of damping~:

\begin{equation}\label{}
\frac{\delta \Gamma}{\omega_0}=\frac{\delta
A_\text{exc}}{A_0}=\frac{\delta F_\text{int}}{k_cA_0},
\end{equation}

\noindent Consequently~:

\begin{equation}\label{}
    \delta E_d=\pi A_0 \delta F_\text{int}
\end{equation}

\subsection{Estimate of $\delta F_\text{int}$}
For large oscillation amplitudes (that is larger than the minimum tip-surface distance, a few angstr\"oms), $F_\text{int}$ is connected to the
so-called normalized frequency shift~\cite{giessibl97a}, $\gamma\equiv\Delta f k_c A_0^{3/2}/f_0$, \emph{via} the equation~\cite{ke99a}
(\emph{cf.} also Ch.16 in ref.[\onlinecite{morita02a}])~:

\begin{equation}\label{}
\gamma(r)\simeq 0.43\sqrt{V_\text{int}(r)F_\text{int}(r)},
\end{equation}

\noindent where $V_\text{int}(r)$ and $F_\text{int}(r)$ are the interaction potential and force, respectively, between the tip and the sample at a
location $r$. The fluctuation in the relative frequency shift $\delta \Delta f/f_0=\delta f/f_0$, that is the cantilever frequency noise, due to a
fluctuation of $F_\text{int}$ is then given by~:

\begin{equation}\label{}
    \frac{\delta f}{f_0}\simeq
    \frac{0.43}{2k_cA_0^{3/2}}\sqrt{\frac{V_\text{int}(r)}{F_\text{int}(r)}}\delta F_\text{int}
\end{equation}

\subsection{Estimate of $\delta f/f_0$}

Y.~Martin \emph{et al.}~[\onlinecite{martin87a}], T.R.~Albrecht
\emph{et al.}~[\onlinecite{albrecht91a}], H.~D\"urig \emph{et
al.}~[\onlinecite{duerig97a}] and F.J.~Giessibl (Ch.2 in
ref.[\onlinecite{morita02a}]) have calculated the thermal limit of
the frequency noise in frequency-modulation technique over a
measurement bandwidth $B$. It is given by~:

\begin{equation}\label{}
    \frac{\delta f}{f_0}=\sqrt{\frac{2k_BTB}{\pi^3k_cA_0^2f_0Q}}
\end{equation}

\noindent Therefore, the dissipated energy due to thermal
fluctuations of the cantilever close to the surface can be
estimated to~:

 \begin{equation}\label{EQU_DELTAED}
    \delta E_d \simeq 4.6 \sqrt{\frac{2k_BTBk_cA_0^3F_\text{int}(r)}{\pi f_0Q V_\text{int}(r)}}
\end{equation}

\noindent The measurement bandwidth $B$ can be estimated out of
the following considerations. As mentioned by F.J.~Giessibl
(\emph{cf.} Ch.2 in ref.[\onlinecite{morita02a}]), $B$ is a
function of the scan speed $v_s$ and the distance $a_0$ between
the features which need to be resolved~:

\begin{equation}\label{EQU_BANDWIDTH}
    B=\frac{v_s}{a_0}
\end{equation}

\noindent For UHV investigations, $a_0$ is of about one atomic lattice constant, that is a few angstr\"oms. At room temperature, due to thermal
drift, atomic scale images are usually recorded at scan speeds of about 6 lines (3 forwards plus 3 backwards) \emph{per} second. Let's consider
for instance a moderate resolution of $6$ pixels \emph{per} atomic period. Then, a line consisting of $256$ pixels should be acquired with a
bandwidth $B=6\times 256 /6=256$~Hz.

Table \ref{TABLE_EDISSVSE0} gives some estimates of the relative dissipated energy due to thermal fluctuations of the cantilever $\delta
E_d/E_{d_0}$ close to the surface in the short-range or pure Van der Waals regimes at various temperatures and for various quality factors. In UHV
at room temperature, our experimental conditions, the minimum dissipated energy which is detectable corresponds to $5\%$ of the intrinsic
dissipated energy of the free cantilever. This corresponds to about 150~meV/cycle with typical conditions for UHV investigations carried out at
room temperature (\emph{cf.} equ.\ref{EQU_ENERGYLOSSINTRINSIC} and discussion below). Besides, as mentioned before, this value is underestimated.
A straightforward consequence is that the strength of apparent dissipation should overcome this limit to be relevant. With a moderate quality
factor in the Van der Waals regime like in high vacuum for instance, the limit drops by almost a factor 3 ($1.8\%$). Thus, apparent dissipation
effects might occur more easily under these conditions~\cite{note_virtual14}. At low temperatures, in the short-range regime the ratio is
$2.67\%$. However, this value is likely still too high because then, the thermal drift being drastically reduced, the measurement bandwidth can be
lowered and apparent dissipation more likely to be measured.

\begin{table}
   \begin{center}
    \begin{tabular}{c c c c c}
       \hline
       \hline
       $Q$                       & Interaction              &   $E_{d_0}$                   & $\delta E_d$                  & $\delta E_d/E_{d_0}$  \\
                                 &   regime                 &   \tiny{(eV/cycle)}       & \tiny{(eV/cycle)}             &                   \\
       \hline
       $5000$   ($298^\circ$K)   & \tiny{VdW + short-range} &  7.69                     &   0.177                       &  $2.3\%$         \\
                                 & \tiny{VdW only}          &  7.69                     &   0.141                       &  $1.8\%$         \\
       $30000$  ($298^\circ$K)   & \tiny{VdW + short-range} &  1.28                     &   $7.25 \times 10^{-2}$       &  $5.7\%$         \\
                                 & \tiny{VdW only}          &  1.28                     &   $5.78 \times 10^{-2}$       &  $4.5\%$         \\
       $500000$ ($4^\circ$K)     & \tiny{VdW + short-range} &  0.077                    &   $2.06 \times 10^{-3}$       &  $2.7\%$         \\
                                 & \tiny{VdW only}          &  0.077                    &   $1.64 \times 10^{-3}$       &  $2.1\%$         \\
       \hline
       \hline
    \end{tabular}
   \end{center}
   \caption{Dissipated energy of the free cantilever $E_{d_0}$ (equ.\ref{EQU_ENERGYLOSSINTRINSIC}) and dissipated energy due to thermal fluctuations of the cantilever close to the surface $\delta E_d$ (equ.\ref{EQU_DELTAED}) for
   various quality factors and temperatures when Van der Waals plus short-range (equ.\ref{EQU_FORCE}) or pure Van der Waals forces (similar equation, with $U_0=0$) are considered. The cantilever parameters are $A_0=7$~nm,
   $f_0=150$~kHz, $k_c=40$~N.m$^{-1}$ and $B=260$~Hz. The parameters of the interaction potential have been taken from ref.[\onlinecite{perez98a}]~: $H=1.865\times 10^{-19}$~J, $R=5$~nm, $U_0=3.641\times 10^{-19}$~J,
 $\lambda=1.2~\AA$, and $r_c=2.357~\AA$. $\delta E_d$ has been estimated at a distance $r$ for which the two interaction regimes are clearly distinct (\emph{cf.} fig.\ref{FIG_FORCEDISTANCE}(a)), $r=5$~$\AA$.
   }\label{TABLE_EDISSVSE0}
\end{table}

\section{Results}\label{SECTION_RESULTS}

\subsection{Validation of the numerical setup}\label{SUBSECTION_VALIDATION}
\begin{figure}
 \includegraphics[height=11cm]{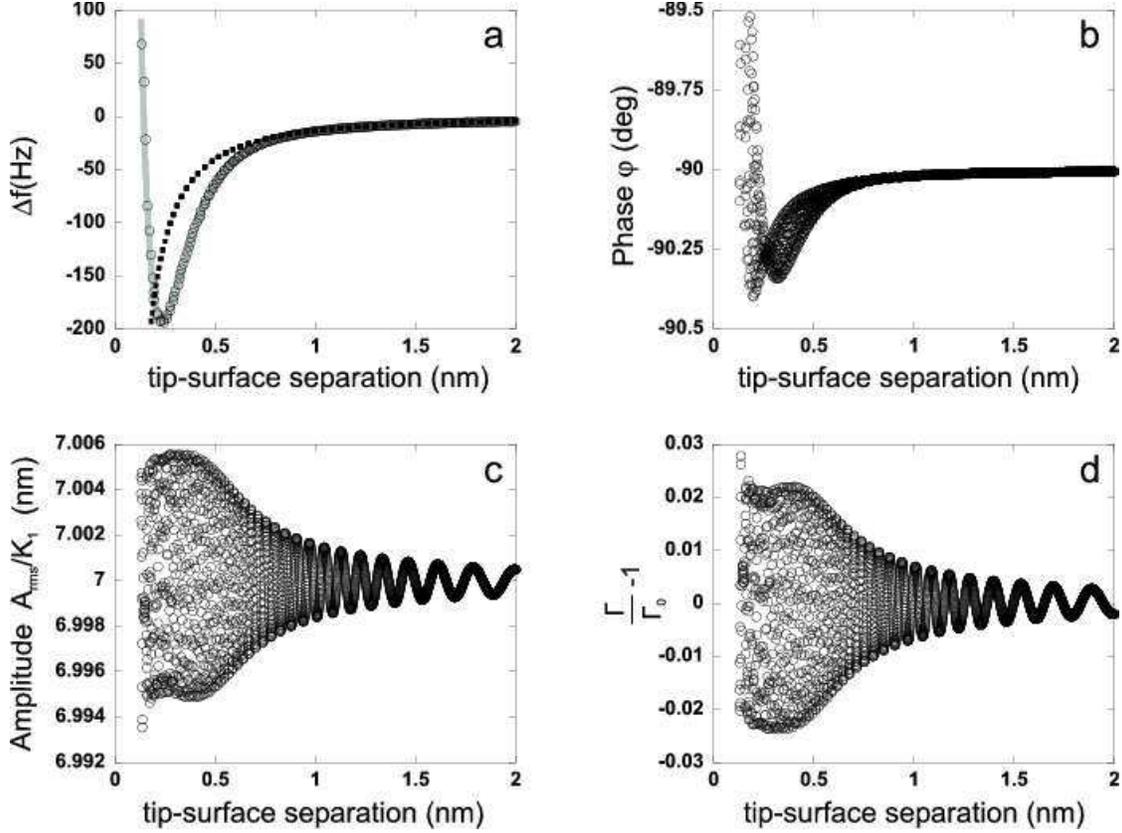}\\
 \caption{Numerical approach curves. The parameters are $A_0^\text{set}=7$~nm, $f_0=150$~kHz, $k_c=30$~N.m$^{-1}$,
 $Q=30000$, therefore $\Gamma_0=31.4$~s$^{-1}$ and $E_{d_0}=0.96$~eV/cycle, $K_1=0.1$~V.nm$^{-1}$, $K_3=1/K_1$~nm.V$^{-1}$, $K_d=1$~V, $K_0=5000$~rad.V$^{-1}$.s$^{-1}$, $K_p^\text{ac}=10^{-3}$,
 $K_i^\text{ac}=10^{-4}$~s$^{-1}$, approach speed $2$~nm.s$^{-1}$. The parameters of the interaction potential have been taken from ref.[\onlinecite{perez98a}]~: $H=1.865\times 10^{-19}$~J, $R=5$~nm, $U_0=3.641\times 10^{-19}$~J,
 $\lambda=1.2~\AA$, and $r_c=2.357~\AA$. Except in (a), the signals
 are monitored at $10$~kHz. (a)-
 Comparison between the $\Delta f$ computed numerically (open circles) and the analytic expression of $\Delta f$ (thick grey line) due to Van der Waals
 and Morse interactions (equ.\ref{EQU_DELTAFANALYT}). The two curves match accurately along the attractive and
 repulsive parts of the interaction potential.
 For clarity reasons, 10 times less samples are displayed compared to plots shown in (b), (c) and (d). The dotted line depicts the analytic $\Delta f$ due
 to a pure Van der Waals potential, thus showing where the short- and long-range interaction regimes are discernable. (b)- Phase lag, $\varphi$. The frequency tracker being engaged, the phase remains constant and equal to $-90$~degrees within deviations limited to $0.3\%$, thus maintaining the cantilever driven on resonance. (c)- Amplitude $A_\text{rms}(t)/K_1$. Since no phase variation occur, the
 amplitude remains constant as well throughout the approach. (d)- Relative damping.}\label{FIG_FORCEDISTANCE}
\end{figure}

Frequency shift \emph{vs.} distance curves obtained from the
simulation have first been compared to the analytic expression of
$\Delta f$ due to Van der Waals and Morse potentials (\emph{cf.}
appendix, section A). The results are shown in
fig.\ref{FIG_FORCEDISTANCE}(a). The parameters chosen to perform
the simulation are consistent with typical parameters used during
experiments performed in UHV. The parameters of the interaction
potential have been taken from ref.[\onlinecite{perez98a}]. They
are representative of the interaction between a silicon tip and a
silicon(111) facet. An excellent agreement is observed between
numerical and analytic curves along the attractive and repulsive
parts of the interaction potential, thus validating the numerical
scheme. The parameters used to perform the calculation are given
in the caption. Let's also notice that the frequency tracker was
engaged. In figs.~\ref{FIG_FORCEDISTANCE}(b), (c) and (d), the
variations of $\varphi$, $A_\text{rms}$ and relative damping,
respectively are reported \emph{vs.} the tip-surface separation.
Phase and amplitude remain almost constant while approaching,
within, however, deviations limited to $0.3\%$ compared to
$-90$~degrees and $A_0^\text{set}=7$~nm, respectively. In the
repulsive part of the potential, steep phase changes occur, but
the amplitude does not dramatically drops, at least up to $\Delta
f= +100$~Hz. Consequently, the relative damping remains constant.

\subsection{Numerical \emph{vs.} real setups}\label{SUBSECTION_VIRTUALVSREAL}
\subsubsection{PLL dynamics}\label{SUBSUBSECTION_PLLDYNAMICS}
\begin{figure}
 \includegraphics[height=6.5cm]{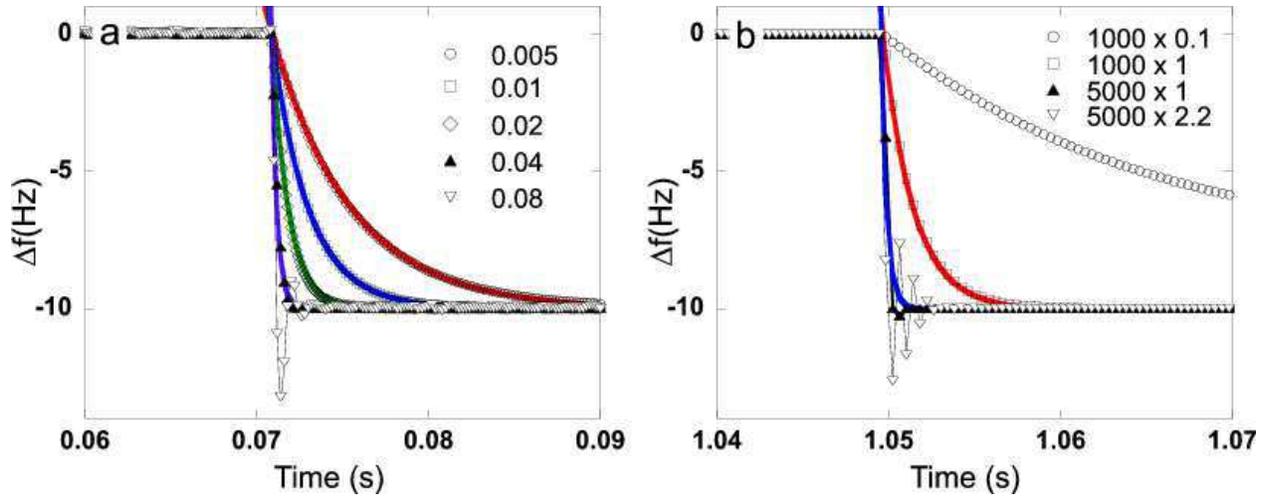}\\
 \caption{(Color online) Step response of the real (a) and of the simulated (b) PLLs to a center frequency
 step of $+10$~Hz at $f_\text{cent}=150$~kHz, resulting in a $\Delta f$ of $-10$~Hz. For this experiment, no interaction between the tip and the surface occurs. The various curves represent
 the experiments carried out for various values of the related
 gains of the PLLs denoted by the symbols. The PLL output is recorded at $10$~kHz. The curves are fitted with a decaying
 exponential (thick continuous lines) out of which the PLL ``locking time" is extracted. They are displayed over similar relative ranges.}\label{FIG_REALVSVIRTUAL}
\end{figure}

The dynamic behaviors of real and simulated PLLs have then been
compared. The experiment consists in locking the PLL onto a
$150$~kHz sinusoidal waveform according to the same procedure than
in section \ref{SUBSUBSECTION_FTRACKER}. The $3$~kHz FIR low pass
filter is used. At a certain moment, a frequency step of $+10$~Hz
is applied to the center frequency, resulting in a shift of
$-10$~Hz ($\omega_0+2\pi \Delta f=\omega_\text{cent}$). The step
response is recorded for various values of the so-called loop gain
(real PLL) and various values of $K_0 K_d$ (simulation). The
variations of $\Delta f$ \emph{vs.} time are fitted with simple
decaying exponential functions, the characteristic time of which
stands for the locking time of the PLL. The results are reported
in figs.\ref{FIG_REALVSVIRTUAL}(a) and (b). A rather long locking
time is noticed for low values of the gains whereas the PLLs lock
faster when the gains become larger. For the latter case, the PLLs
can operate up to the limit of the locking range as shown by the
oscillations.

\begin{figure}
 \includegraphics[height=8cm]{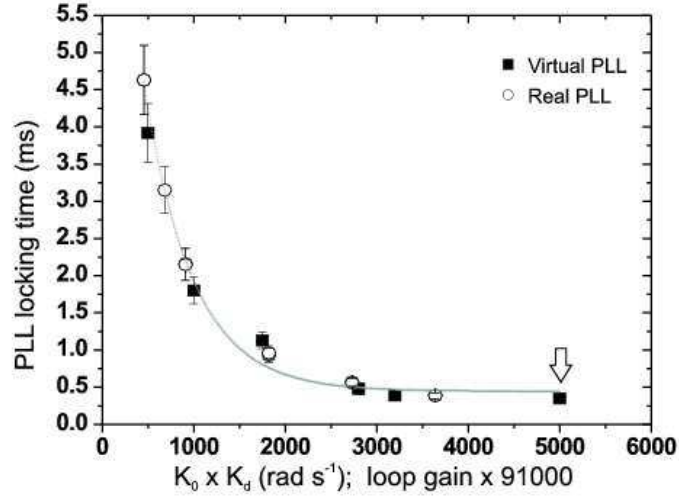}\\
 \caption{Locking time of the simulated (filled squares) and real (empty circles) PLL \emph{vs.} $K_0 K_d$ and loop gain$\times 91000$. The locking times are
 obtained from the related step response curves (figs.\ref{FIG_REALVSVIRTUAL}(b) and (a), respectively). The error bars depict the uncertainty on the
 fitted value of the locking time ($\pm10\%$). The arrow indicates the value of the loop gain used experimentally which corresponds to an optimum behavior of the PLL and a
 related locking time of about 0.35~ms. The curve is given as guide eyes.}\label{FIG_REALVSVIRTUAL_LOCKINGTIME}
\end{figure}

The locking time deduced from each fit is plotted as a function of
the gains of both PLLs. In order to make the curves comparable,
the loop gain must be rescaled by an arbitrary constant which
depends on the electronics. The best agreement between the curves
was achieved with $91 \times 10^3$ (\emph{cf.}
fig.\ref{FIG_REALVSVIRTUAL_LOCKINGTIME}). A single master curve of
the PLL dynamics can thus be extracted. The rather good agreement
between the two curves provides evidence that the simulation
reasonably describes the real component, at least within the
locking range. For values of the gains up to
$K_0K_d=6000$~rad.s$^{-1}$, the PLL is stable and able to track
frequency changes within the locking range around $150$~kHz. Above
6000~rad.s$^{-1}$, the PLL introduces overshoot in the output
waveform while attempting to lock the input signal. For higher
gains, the PLL is not able to properly track the input signal,
even though its frequency is within the locking range. The border
is reached for $K_0>10^4$~rad.V$^{-1}$.s$^{-1}$, in good agreement
with the value expected from R.Best's criterion (\emph{cf.}
discussion in section \ref{SUBSUBSECTION_FDEMODULATION}). The
arrow in fig.\ref{FIG_REALVSVIRTUAL_LOCKINGTIME} indicates the
usual loop gain value which is chosen to perform the experiments
using the $3$~kHz low pass filter, corresponding to
$K_0K_d=5000$~rad.s$^{-1}$. The related locking time of the PLL is
then $\simeq 0.35$~ms. For those values of gains, the locking
range is about $\pm 400$~Hz.

\subsubsection{APIC dynamics}\label{SUBSUBSECTION_APICDYNAMICS}
\begin{figure}
 \includegraphics[height=6.5cm]{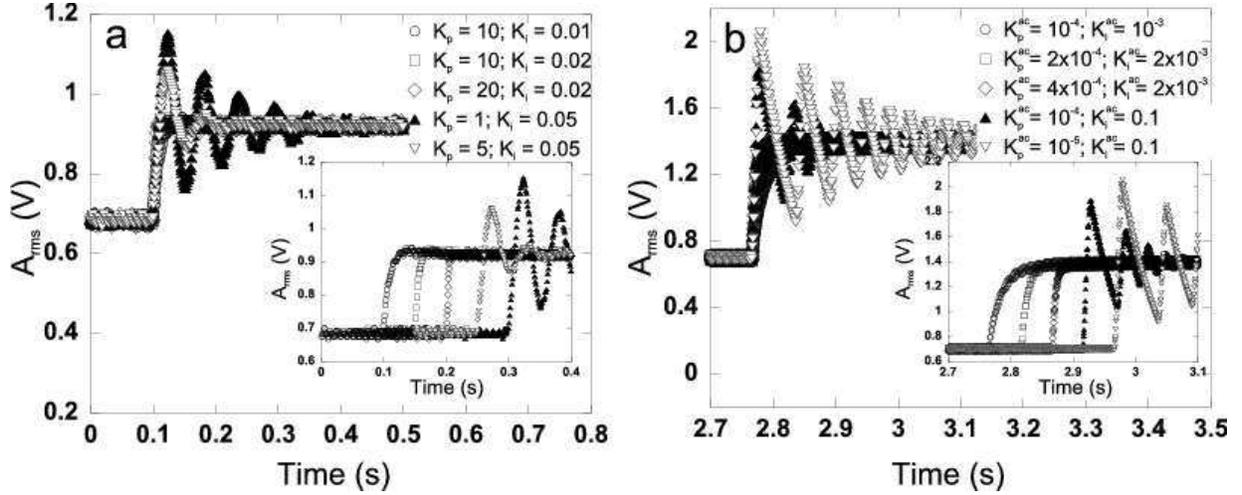}\\
 \caption{Step response of the real (a) and of the simulated (b) APIC to a $A_0^\text{set}$ step. To perform the experiments, the related PLLs
 are engaged (frequency trackers as well). The $3$~kHz FIR low pass filter has been used. No interaction between the tip
 and the surface occurs. The cantilever properties are $f_0=157514.6$~Hz and $Q=36000$. To perform the calculation,
 since the cantilever stiffness was not accurately known, we have arbitrarily chosen $k_c=30$~N.m$^{-1}$, in reasonable
 agreement with manufacturer's datasheet. The other numerical parameters are similar to those given in fig.\ref{FIG_FORCEDISTANCE}. The curves
 depict the experiments carried out for various values of $K_p^\text{ac}$ and $K_i^\text{ac}$ gains. Three behaviors are observed~: overcritically damped responses without overshoot, critically
 damped responses with a slight overshoot and undercritically damped responses with an oscillating behavior. The
 insets show them for couple of curves which have been
 arbitrarily shifted along the time axis, but note that the relative ranges are similar.}\label{FIG_APICSTEPRESPONSE}
\end{figure}

\begin{figure}
 \includegraphics[height=8cm]{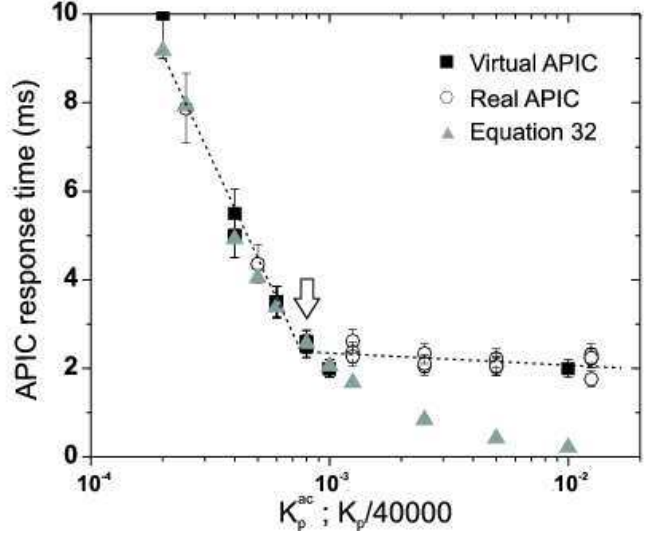}\\
 \caption{Response time of the APIC \emph{vs.} $K_p^\text{ac}$ of the simulated setup and the rescaled $K_p$ gain of the
 real controller. The best agreement between the curves is obtained with $K_p/40000$. The two curves match with a reasonable agreement and exhibit two domains~: first the response time
 decreases when increasing $K_p^\text{ac}$ and then a saturation is reached corresponding to $t_\text{resp}\simeq 2$~ms. The dotted line is
 given as guide eyes. Such an analysis can be performed assuming that the step response is governed
 by a single time constant, thus restricting the analysis to curves which exhibit an almost critically damped behavior (\emph{cf.} text).
 The triangles depict the trace of the function $t_\text{resp}$ (equ. \ref{EQU_TIMERESPONSEAPIC}) with $f_0=157514.6$~Hz,
 $Q=36000$, $K_1=0.1$~V.nm$^{-1}$ and $K_3=1/K_1$~nm.V$^{-1}$. The arrow denotes the value of the gain used to perform the scan
 lines (\emph{cf.} section \ref{SECTION_RESULTS}). Beyond $K_p^\text{ac}=10^{-3}$, a noticeable discrepancy between the
 theoretical model and the experiments is observed which might be due to the contribution of the RMS-to-DC converter.}\label{FIG_APICTIMERESPONSE}
\end{figure}

In order to extract a typical time constant of the component,
similar experiments have been carried out with the APICs. The
cantilever remaining far from the surface, a step is applied to
the setpoint amplitude $A_0^\text{set}$ resulting in an abrupt
change in $A_\text{rms}$ upon gains. The results are reported in
fig.\ref{FIG_APICSTEPRESPONSE}(a, real setup) and (b, simulated
setup). The curves exhibit over- (no overshoot at all) under-
(oscillating behavior) or critically damped (single overshoot)
behaviors upon chosen gains. So as to extract the APIC response
time, we focus at curves which exhibit a single time constant,
that is curves for which a weak overcritically or a critically
damped response is observed (\emph{cf.} insets in
figs.\ref{FIG_APICSTEPRESPONSE}(a) and (b)). This is motivated by
the controller response which is then the fastest, while
preserving an overall stable behavior. The changes in
$A_\text{rms}$ are fitted with decaying exponentials functions and
the related characteristic time is extracted. The variation of the
so-called response time of the controller ($t_\text{resp}$)
\emph{vs.} gains is reported in fig.\ref{FIG_APICTIMERESPONSE}.
The restriction to curves exhibiting a single time constant is
similar to restricting the analysis to a single gain of the
controller~\cite{note_virtual18}. Thus, a single master curve
which describes the dynamics of both APICs can be extracted as
well. In fig.\ref{FIG_APICTIMERESPONSE}, the $K_p$ gain of the
real controller has been rescaled to make it matching
$K_p^\text{ac}$ (the best rescaling factor is 1/40000).
$t_\text{resp}$ decreases as $K_p^\text{ac}$ increases (being
given a single $K_i^\text{ac}$ \emph{per} $K_p^\text{ac}$).
However, the controller is limited to an optimum $t_\text{resp}$
of about $2$~ms as shown by the plateau reached for
$K_p^\text{ac}\simeq 10^{-3}$~[\onlinecite{note_virtual01}] (arrow
in fig.\ref{FIG_APICTIMERESPONSE}).

So far, the origin of the saturation remains unclear.
Nevertheless, a brief analysis of the response function of the
controller to a step wherein the contribution of the RMS-to-DC
converter is neglected (\emph{cf.} appendix, section B) emphasizes
that the dynamic behavior can reasonably be predicted (triangles
in fig.\ref{FIG_APICTIMERESPONSE}) up to $2$~ms. The best
agreement between the experimental results and the model is found
when considering the weak overcritically damped regime, namely~:

\begin{equation}\label{EQU_TIMERESPONSEAPIC}
t_\text{resp}\simeq
\frac{1}{c+\sqrt{c^2-\frac{\omega_0}{2}K_1K_3K_i^\text{ac}}},
\end{equation}

\noindent with~:

\begin{equation}\label{EQU_C}
    c=\frac{\omega_0}{4}\left( \frac{1}{Q}+K_1K_3K_p^\text{ac}\right)
\end{equation}

\noindent The origin of the saturation might thus be attributed to
the contribution of the RMS-to-DC converter.

As expected, the shortest APIC response time is approximately $6$
times longer than the optimal PLL locking time, $\simeq 0.35$~ms.
Thus the PLL should track frequency changes much faster than
amplitude changes. Therefore, with PLL gains insuring a locking
time much shorter than the APIC one, the two blocks can be
considered as operating separately. Then, no amplitude changes
which would be the consequence of a bad tracking of the resonance
frequency can occur.

It might be objected that the experiments and the analysis, despite consistent, have been performed without considering the tip-sample
interaction. Regarding the PLL, the way the dynamics is affected when the tip is close to the surface has not yet been investigated. But regarding
the amplitude controller, Couturier \emph{et al.}~[\onlinecite{couturier03a}] have reported a theoretical analysis of the controller stability
upon the gains and the strength of the non-linear interaction in the self-excitation scheme. The analysis stresses that the stability domain of
the controller shrinks when the contribution of the non-linear interaction (pure Van der Waals) increases. Thus, a couple
($K_p^\text{ac}$;$K_i^\text{ac}$) initially inside the stability domain might correspond to an unstable behavior of the controller close to the
surface, thus introducing apparent dissipation~\cite{couturier03a}. Nevertheless, considering their parameters with a tip-surface distance ranging
from infinity down to $0.8$~$\AA$ (corresponding to $\Delta f\lesssim-250$~Hz), that is very close to the surface for operating in
nc-AFM~\cite{note_virtual07}, the stability domain weakly shrinks~\cite{note_virtual06}. A similar analysis for the PLL-excitation scheme is still
lacking and should be performed for quantitative comparison and discussion. But comparing their analysis to the tip-surface distances and
frequency shifts which are being used in this work, we believe that the contribution of the non-linear interaction to the APIC dynamics remains
weak and thus would not change drastically its coupling to the PLL. This point is strengthened by the results given in the following section
(\ref{SUBSUBSECTION_PLLGAINS}).

\subsection{Apparent dissipation}\label{SUBSECTION_APPARENTDAMPING}
\subsubsection{Contribution of the PLL gains}\label{SUBSUBSECTION_PLLGAINS}
\begin{figure}
 \includegraphics[height=11cm]{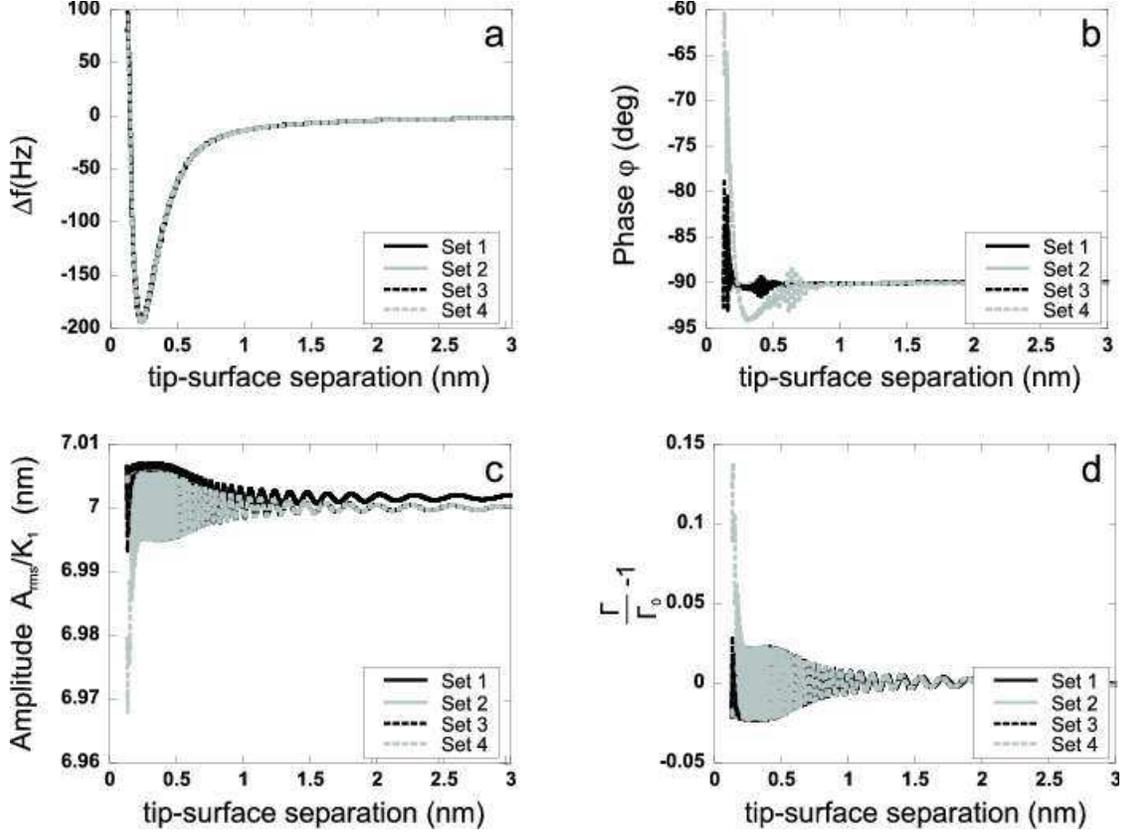}\\
 \caption{Approach \emph{vs.} distance curves for various sets ($1$ to $4$) of $K_0 K_d$ of the Pll gains,
 namely~: $11000$ (continuous black line), $5000$ (continuous grey line), $1000$ (dotted black line) and
 $100$~rad.s$^{-1}$ (dotted grey line), corresponding to locking times of 0.2, 0.35, 1.8 and $>4$~ms,
 respectively. (a)- $\Delta f$, the curves are all matching each other. (b)- Phase lag $\varphi$, (c)- amplitude $A_\text{rms}(t)/K_1$
 and (d)- relative damping. Except $K_0K_d$, the parameters are similar to those given in fig.\ref{FIG_FORCEDISTANCE}, in particular $K_p^\text{ac}=10^{-3}$
 and $K_i^\text{ac}=10^{-4}$~s$^{-1}$ corresponding to $t_\text{resp}\simeq 2$~ms.
When the PLL locking time is larger than $t_\text{resp}$ (set $4$,
$K_0 K_d=100$), the resonance is not properly locked, which
induces a phase shift. Consequently, the amplitude decreases and
the damping increases.}\label{FIG_PLLGAINS}
\end{figure}

Section \ref{SUBSUBSECTION_PLLDYNAMICS} has proved that the PLL
gains were controlling the PLL locking time. Within the locking
range, the higher $K_0K_d$, the faster the PLL. The test performed
here is to compute approach curves for various values of $K_0K_d$.
Except $K_0K_d$, the parameters are similar to those given in
fig.\ref{FIG_FORCEDISTANCE}. In particular, $Q=30000$,
$K_p^\text{ac}=10^{-3}$ and $K_i^\text{ac}=10^{-4}$~s$^{-1}$,
corresponding to $t_\text{resp}=2$~ms. $4$ sets of $K_0 K_d$ have
been used, namely~: $11000$, $5000$, $1000$ and
$100$~rad.s$^{-1}$, corresponding to locking times of $\simeq
0.2$~ms, $0.35$~ms, $1.8$~ms and $>4$~ms, respectively. Note that
the two later values are almost similar or larger, respectively,
than $t_\text{resp}$. The $3$~kHz FIR low pass filter has been
used and the frequency tracker has been engaged. The results are
reported in fig.\ref{FIG_PLLGAINS}.

With the four sets of data, no effect on the frequency shift is
observed. For $K_0 K_d=11000$ and $5000$~rad.s$^{-1}$, changes in
phase, amplitude and damping are noticeably similar. The phase and
the amplitude remain constant and subsequently, no damping occurs.
On the opposite, for $K_0 K_d=1000$ and $100$~rad.s$^{-1}$, that
is for a PLL locking time of about or larger than the APIC one,
the changes are more pronounced. With $K_0 K_d=1000$~rad.s$^{-1}$
(set 3), the phase strongly varies along the repulsive part of the
interaction, from $-93$ to $-79$~degrees. Consequently, the
amplitude starts dropping and the damping increases. This trend is
more pronounced with $K_0 K_d=100$~rad.s$^{-1}$, for which the
phase in the repulsive regime reaches $-60$~degrees, requiring the
APIC to produce $\simeq 14\%$ more excitation. According to the
discussion put forward in section \ref{SECTION_MINDISSDETECT},
such an effect is expected to be detected if it would occur.

Therefore, if the PLL does not lock the incoming signal fast
enough, that is for locking times of about or larger than $1$~ms,
corresponding approximately to $t_\text{resp}/2$, an undesirable
phase shifted signal is produced, resulting in an amplitude
decrease and producing significant apparent dissipation.

It's peculiar to notice that, when the cantilever is driven out of
resonance as shown with the phase changes with
$K_0K_d=100$~rad.s$^{-1}$, no abnormal frequency shift occurs. As
a matter of fact, close to the resonance, the phase changes of the
free cantilever scale as, to first order~: $\varphi \overset{u
\rightarrow 1}{=}\pi /2 -2Q(u-1)$, where $u=f/f_0$. Considering
$f_0=150$~kHz and $\delta \varphi=\pm 30$~degrees, then $\delta
f=f_0\delta \varphi/2Q=\pm 1.3$~Hz, which is not visible in
fig.\ref{FIG_PLLGAINS}(a). Similar effects have been reported by
H.H\"olscher \emph{et al.}~[\onlinecite{hoelscher01a}]. This
effect should be more (less) pronounced with low (high) $Q$ values
($\pm 8$~Hz or $\pm 0.08$~Hz with $Q=5000$ or $500000$,
respectively) and the apparent dissipation higher (lower).

\subsubsection{Contribution of the frequency tracker}\label{SUBSUBSECTION_FREQUTRACK}
\begin{figure}
 \includegraphics[height=11cm]{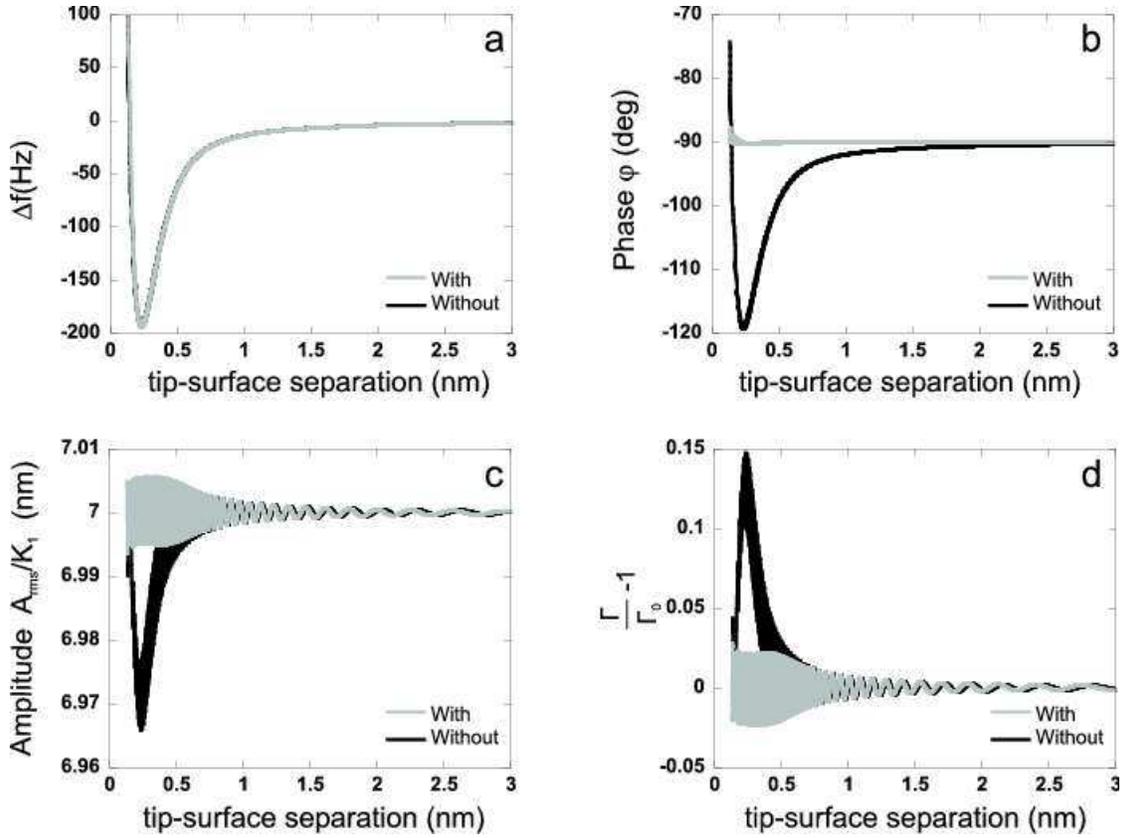}\\
 \caption{Approach \emph{vs.} distance curves upon the frequency tracker of the PLL is engaged or not (grey or black lines,
 respectively).(a)- Frequency shift $\Delta f$, (b)- phase lag $\varphi$, (c)- amplitude $A_\text{rms}(t)/K_1$ and (d)- relative damping.
 The parameters are similar to those given in the caption of fig.\ref{FIG_FORCEDISTANCE}. When not engaged, the phase continuously drifts during the approach due to the increase of the attractive interaction meaning that the
 oscillator is not driven on resonance. Subsequently, the amplitude drops and the APIC strives to keep it constant by increasing
 the excitation. $15\%$ more excitation is thus produced, that is above thermal noise (\emph{cf.} section \ref{SECTION_MINDISSDETECT}).}\label{FIG_PLLFREQUTRACK}
\end{figure}

As mentioned in paragraph \ref{SUBSUBSECTION_FTRACKER}, the
frequency tracker updates $\omega_\text{cent}$ with the goal to
prevent the phase due to the frequency shift,
$\varphi_\text{pll}$, be added to the NCO output. Figure
\ref{FIG_PLLFREQUTRACK} reports two approach curves computed upon
the frequency tracker is engaged or not. When it is disengaged,
the phase continuously decreases along the attractive part of the
interaction potential, as expected from equ.\ref{EQU_PHIPLL}. At a
tip-sample separation corresponding to the minimum of the
interaction potential, $r=2.35$~$\AA$, the phase reaches
$-120$~degrees, meaning that the oscillator is then seriously
driven out of resonance. Following the phase change, the amplitude
continuously decreases and the damping strives to compensate the
amplitude reduction, thus reaching $15\%$ of the intrinsic damping
of the free oscillator. Here again, such an effect should be
measurable. When the tip is further approached towards the
surface, the repulsive regime makes the frequency shift increasing
and so does $\varphi_\text{pll}$. The amplitude increases back to
reach $A_0^\text{set}$ and the damping is obviously reduced. An
amplitude growth is not expected in the repulsive region of the
interaction potential, but it is the consequence of the bad
tracking of the resonance frequency. On the other hand, when the
tracker is engaged, as already mentioned, the phase remains
constant and no apparent dissipation occurs.

\subsection{Scan lines}\label{SUBSECTION_SCANS}
In addition to the analysis of the time constant of the various blocks, it is important to focus at variables changes when the tip is scanned
along a surface. Two types of surfaces have been investigated~: 1- a sinusoidally corrugated surface with a spatial wavelength of $6.6~\AA$ and a
corrugation of $\pm 0.1$~nm, consistently with the lattice constant of KBr, a sample regularly used in the group, and 2- a surface with two
opposite steps with a step height of $3.3~\AA$. The steps are built out of $\arctan$ functions and spread out laterally over $5~\AA$. The upper
terrace spreads out over $3$~nm (\emph{cf.} insets in fig.\ref{FIG_SCANAPIC}).

The results shown here have all been obtained by $\Delta f$
regulation. The scan lines have been initiated from the approach
curve shown in fig.\ref{FIG_FORCEDISTANCE} with $\Delta
f_\text{set}=-60$~Hz, corresponding to an initial tip-surface
separation of about $5~\AA$, that is in the short-range regime
(\emph{cf.} fig.\ref{FIG_FORCEDISTANCE}(a)). The gains of the
distance controller have been chosen in order to insure a
critically damped response of the controller to a step of $-1$~Hz
when $\Delta f_\text{set}$ is reached, namely
$K_p^\text{dc}=2\times 10^{-3}$~nm.Hz$^{-1}$ and
$K_i^\text{dc}=2$~nm.Hz$^{-1}$.s$^{-1}$. For all of the following
curves, the frequency tracker has been engaged. Three sets of
parameters have been varied~: the PLL gains, the scan speed and
the APIC gains.

\subsubsection{Contribution of the PLL gains}
Paragraph \ref{SUBSUBSECTION_PLLDYNAMICS} has proven how $K_0$ and
$K_d$ gains were controlling the PLL locking time. Figure
\ref{FIG_SCANPLL} shows scan lines computed for three values of
$K_0 K_d$, namely~: $100$, $1000$ and $5000$~rad.s$^{-1}$,
corresponding to locking times $>4$, $1.8$ and $0.35$~ms,
respectively. The unchanged parameters are~: scan speed
$=7$~nm.s$^{-1}$, $K_p^\text{ac}=10^{-3}$ and
$K_i^\text{ac}=10^{-4}$~s$^{-1}$. The latter gains correspond to
$t_\text{resp}\simeq 2$~ms (arrow in
fig.\ref{FIG_APICTIMERESPONSE}). Each signal, namely topography,
$\Delta f$, $\varphi$, $A_\text{rms}$ and relative damping,
consists of 256 samples. The topography signal follows accurately
the surface corrugation. No contribution due to the gains is
revealed. $\Delta f$ is modulated around $\Delta
f_\text{set}=-60$~Hz, with an amplitude ranging from $\pm2$ to
$\pm3$~Hz upon $K_0K_d$. The accuracy of the distance control is
then of about $97\%$. Note also that the nonlinear interaction
makes the modulation asymmetric around $\Delta f_\text{set}$ and
the maxima are mismatched compared to the maxima of the surface.
However, the mismatch does not depend on $K_0K_d$. If the PLL is
slow, a rather important phase lag is observed, ranging from $-65$
to $-110$ degrees, resulting in small amplitude changes. Here
also, the asymmetry around $-90$~degrees is manifest and it's
interesting, despite expected, to notice the doubling in the
periodicity of the amplitude fluctuation. The related relative
apparent damping fluctuates accordingly, reaching about
$12\%\Gamma_0$, which should be experimentally detectable.

\subsubsection{Contribution of the scan speed}
Five scan speeds ranging from $1$ to $20$~nm.s$^{-1}$ have been
used, accordingly to typical experimental values for such a scan
size. The unchanged parameters are~: $K_0 K_d=5000$~rad.s$^{-1}$,
$K_p^\text{ac}=10^{-3}$ and $K_i^\text{ac}=10^{-4}$~s$^{-1}$. The
surface with opposite steps has been used and each signal consists
of 256 samples. The results are given in fig.\ref{FIG_SCANSPEED}.
At high speed, the topography channel starts being distorted as a
consequence of a bad distance regulation as shown by the large
$\Delta f$ variations. The phase varies accordingly, but within a
narrower domain. Therefore neither relevant variations of
amplitude nor of relative damping ($\pm 2\%$ only) are revealed.

\subsubsection{Contribution of the APIC gains}
Similar experiments have been carried out by varying the APIC
gains. Thirteen sets of $K_p^\text{ac}$ and $K_i^\text{ac}$ values
have been used over 2 orders of magnitude for each gain. The
unchanged parameters are~: scan speed $= 5$~nm.s$^{-1}$ and $K_0
K_d=5000$~rad.s$^{-1}$. The surface with opposite steps has been
used. For those curves, each signal consists of 1024 samples. The
sets of gains have been chosen such that the response time of the
controller is varied from $2$ to $20$~ms, according to
fig.\ref{FIG_APICTIMERESPONSE}. The results are reported in
fig.\ref{FIG_SCANAPIC}. Here again, the topography signal
accurately follows the corrugation. In particular, no unwanted
overshoot is observed at the step edge despite $\Delta f$ varies
significantly. A small phase variation is observed at the step
edge. For the latter three channels, no dependence is observed
upon the gains such that the curves match each other. The small
phase variation induces a tiny amplitude change, barely visible in
the inset of fig.\ref{FIG_SCANAPIC}(d), but the overall
fluctuations are weak, which corresponds to relative damping
fluctuations of $\pm 3\%$ (worse case,
$t_\text{resp}\approx20$~ms). Following discussion of section
\ref{SECTION_MINDISSDETECT}, this value is below the threshold
limit of thermal noise, at least for experiments carried out in
UHV and at room temperature. Upon gains, a small spatial shift is
observed in the amplitude or in the relative damping signals up to
a maximum value of $0.1$~$\AA$.

\section{Discussion}
The above analysis stresses five important results~:

\begin{itemize}
    \item The PLL dynamics plays a major role in the occurrence of relevant apparent
    damping if the locking time is about or larger than 1~ms, that is only twice faster than the APIC optimum response time.
    By ``relevant damping", we mean, on the base of the discussion given in section \ref{SECTION_MINDISSDETECT}, a
    damping which would be detectable experimentally upon operating conditions (\emph{cf.} table \ref{TABLE_EDISSVSE0}).
    \item The frequency tracker, the aim of which is to update the PLL center frequency to make it matching the
    actual resonance frequency, plays also a major role in the occurrence of apparent
    damping. It has to be mandatorily engaged when performing
    approach-curves, otherwise unwanted additional phase shift due to the PLL occurs and the cantilever is then driven off resonance.
    \item The PLL optimal locking time is about $0.35$~ms that is 6 times shorter than the
    shortest APIC response time of the free cantilever. Therefore the resonance condition is expected to be always properly
    maintained. Consequently, when the PLL operates properly, no amplitude changes due to a bad tracking of the
    resonance frequency are expected to occur. If they would, this should rather be the consequence of the
    APIC and/or the DPIC dynamics.
    \item  The APIC response time seems to be limited to $t_\text{resp}\simeq 2$~ms due to the RMS-to-DC converter.
    There is \emph{a priori} no fundamental restriction to the APIC response time, as shown by fig.\ref{FIG_APICTIMERESPONSE}.
    However, it must be stressed that if the APIC is made faster, the PLL should be made faster accordingly.
    \item  A weak contribution of the APIC to apparent dissipation is observed. Although spatial shift and
     apparent dissipation can conditionally be generated, the overall strength of the effect
    remains weak and should hardly be measurable for UHV investigations at room temperature.
\end{itemize}

In order to compare our results to other works, the contribution
of the noise (equ.\ref{EQU_DELTAED}) to the dissipation has been
estimated with the parameters given by Gauthier \emph{et
al.}~[\onlinecite{gauthier02a}]. The maximum of relative apparent
damping they report is about $5\%$ (\emph{cf.} fig.4 in the above
reference, curve $\sharp$6), corresponding to a scanning speed of
about $90~\AA$.s$^{-1}$. The spatial wavelength of their surface
model being $8\AA$, the bandwidth of the corresponding measurement
is $B\simeq 11$~Hz (\emph{cf.} equ.\ref{EQU_BANDWIDTH}). The
non-linear interaction is depicted by a Rydberg function with
$U_0=4\times 10^{-11}$~J, $\lambda=0.599~\AA$ and $r_c=3~\AA$.
With $A_0=1.5$~nm, $f_0=159154$~Hz, $k_c=26$~N.m$^{-1}$, $Q=24000$
($E_{d_0}=4.8\times 10^{-2}$~eV/cycle), we get $\delta
E_d/E_{d_0}\simeq 2.3\%$ at a distance~\cite{note_virtual09}
$s=4~\AA$. Therefore, due to the use of a low
amplitude~\cite{note_virtual15}, the strength of the effect would
not be balanced by the noise and could be detectable
experimentally. On the opposite, considering smaller scan speeds,
the maximum of apparent dissipation decreases down to $2\%$. Such
effects become then unlikely to be observed experimentally.

\begin{figure}
 \includegraphics[width=14cm]{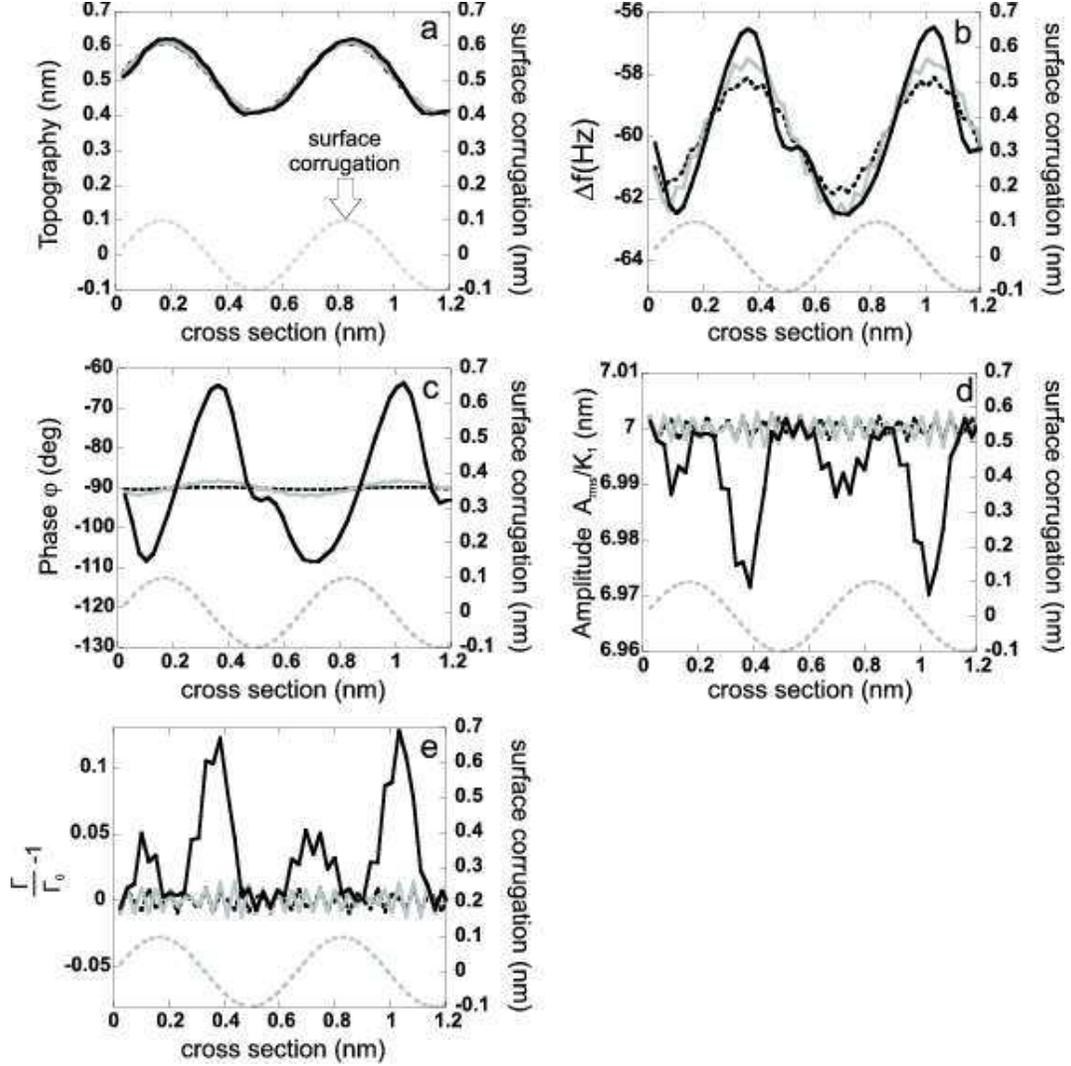}\\
 \caption{Calculated cross-section of a sinusoidally corrugated surface for various PLL gains, namely $K_0K_d=100$~rad.s$^{-1}$ (continuous black line),
  $1000$~rad.s$^{-1}$ (continuous grey line) and $5000$~rad.s$^{-1}$ (dotted black line). The scan lines have been initiated from the approach
  curve shown on fig.\ref{FIG_FORCEDISTANCE} by $\Delta f$ regulation using $\Delta f_\text{set}=-60$~Hz. The lateral scan speed is $7$~nm.s$^{-1}$ and
  the section consists of 256 samples. $K_p^\text{dc}=2\times 10^{-3}$~nm.Hz$^{-1}$ and $K_i^\text{dc}=2$~nm.Hz$^{-1}$.s$^{-1}$,
  corresponding to a critically damped response of the controller to a frequency step of $-1$~Hz. (a)- Topography.
  (b)- Frequency shift. (c)- Phase lag $\varphi$. (d)- Amplitude $A_\text{rms}/K_1$ and (e)- relative damping. No noticeable effect
  is revealed on the topography. The apparent dissipation remains below the thermal noise (\emph{cf.} section \ref{SECTION_MINDISSDETECT}), except if the PLL is slow ($K_0K_d=100$~rad.s$^{-1}$, $12\%$).} \label{FIG_SCANPLL}
\end{figure}

\begin{figure}
 \includegraphics[width=14cm]{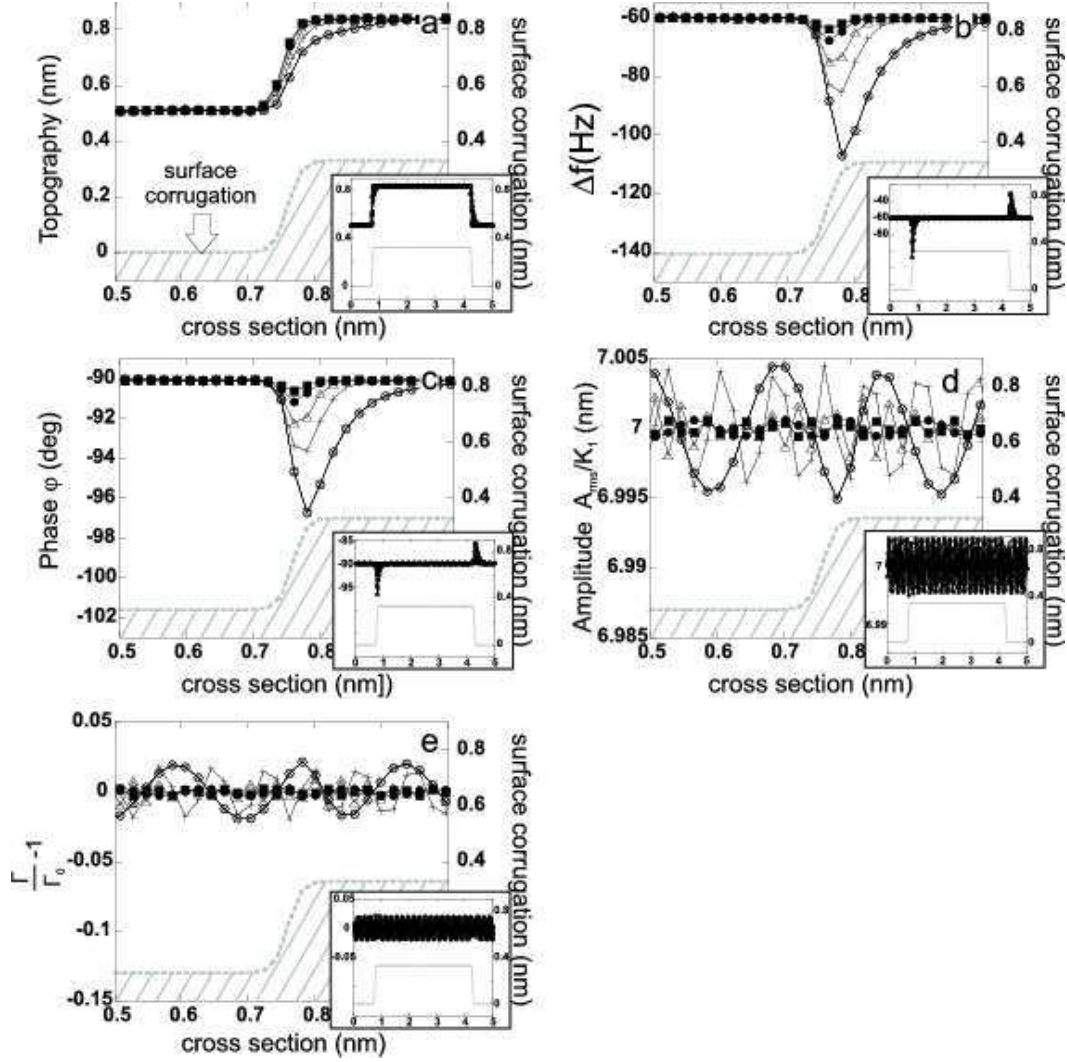}\\
 \caption{Calculated cross-section of a surface with two opposite steps for various scanning
 speeds, namely 1 ($\blacksquare$), 2 ($\bullet$), 5 ($\vartriangle$), 10 ($+$) and 20~nm.s$^{-1}$ ($\circ$). For clarity reasons, the right hand side step region has been magnified.
 The insets show the whole section. The scan lines have been initiated upon the same conditions than in fig.\ref{FIG_SCANPLL} with $K_0K_d=5000$~rad.s$^{-1}$, $K_p^\text{ac}=10^{-3}$ and
 $K_i^{ac}=10^{-4}$~s$^{-1}$. The high speeds require to reduce the number of samples \emph{per} line to 256. The gains of the distance controller are similar to those given in
 fig.\ref{FIG_SCANPLL}. (a)- Topography. (b)- Frequency shift. (c)- Phase lag $\varphi$. (d)- Amplitude $A_\text{rms}/K_1$ and (e)- relative damping.
 At high scan speeds, the topography is slightly distorted, but the overall damping remains weak and below the thermal noise (\emph{cf.} section \ref{SECTION_MINDISSDETECT}).}\label{FIG_SCANSPEED}
\end{figure}

\begin{figure}
 \includegraphics[width=14cm]{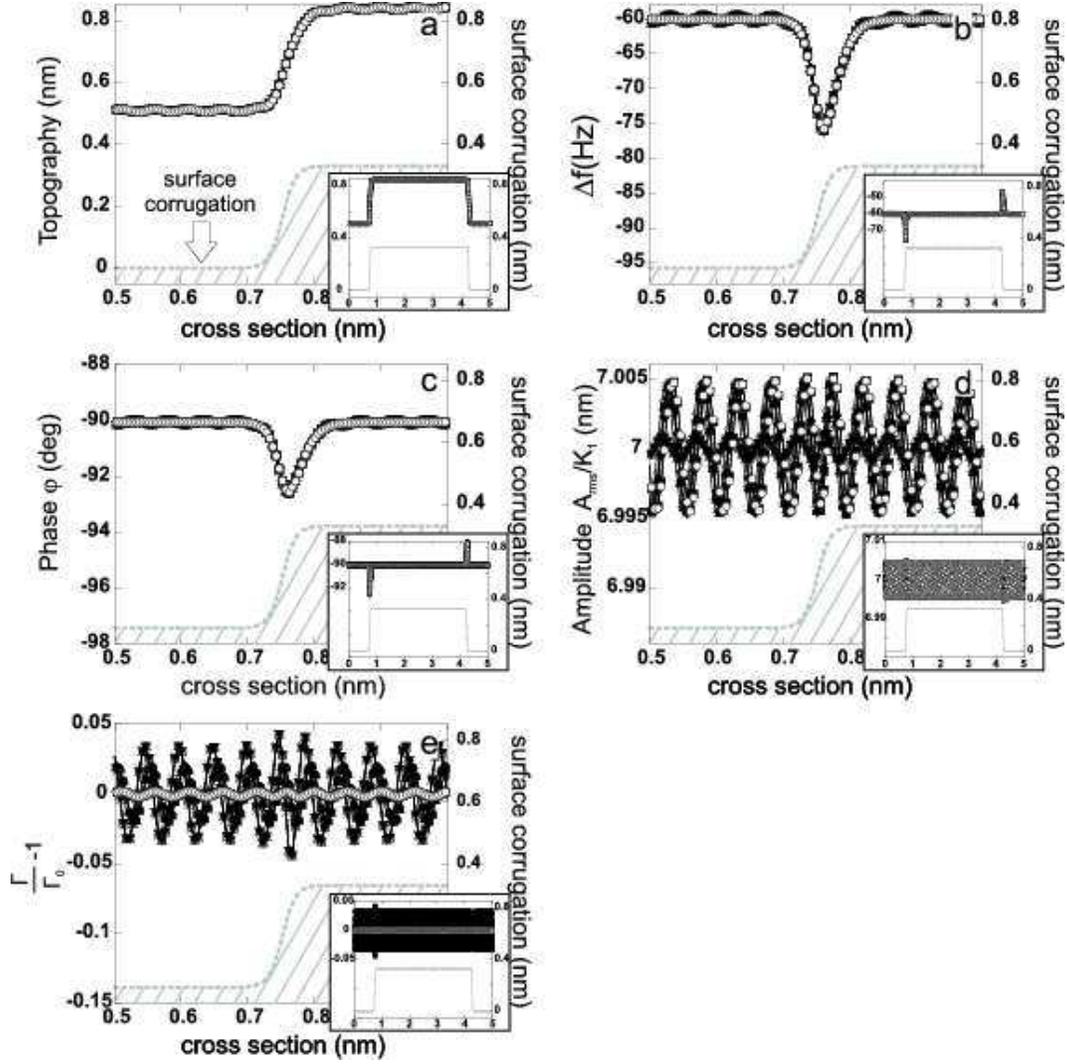}\\
 \caption{Calculated cross-section of a surface with two opposite steps for various sets of the APIC gains,
 namely $(K_p^\text{ac}=10^{-2};K_i^\text{ac}=10^{-3}$~s$^{-1})=\circ$, $(10^{-3}; 10^{-3})=\diamond$,
 $(10^{-2}; 10^{-2})= \times$, $(10^{-3}; 10^{-2})= +$, $(10^{-2}; 10^{-4})=
 \vartriangle$, $(10^{-3}; 10^{-4})= \bullet$, $(10^{-4}; 10^{-4})= \blacksquare$, $(10^{-4}; 10^{-3})= \blacklozenge$,
 $(10^{-3}; 10^{-5})= \blacktriangle$, $(10^{-2}; 10^{-5})= \blacktriangledown$ and $(10^{-4}; 10^{-5})= \odot$. The scan lines have been initiated upon the same conditions than in fig.\ref{FIG_SCANPLL} with $K_0K_d=5000$~rad.s$^{-1}$. The lateral scan
 speed is $5$~nm.s$^{-1}$ and the section consists of 1024 samples. The gains of the distance controller are similar to those given in fig.\ref{FIG_SCANPLL}. (a)- Topography.
  (b)- Frequency shift. (c)- Phase lag $\varphi$. (d)- Amplitude $A_\text{rms}/K_1$ and (e)- relative damping. The APIC gains have a negligible effect on the
  topography. A weak apparent damping ($3\%\Gamma_0$, that is not experimentally relevant) is revealed at the step if the APIC is slow, corresponding to $t_\text{resp}=20$~ms.}\label{FIG_SCANAPIC}
\end{figure}

Finally, let's note that the contribution of the third controller,
the DPIC, has not been assessed in this work. Nevertheless, as
mentioned by H.~Hug and A.~Baratoff~[\onlinecite{note_virtual08}],
in order to minimize feedback errors and resulting image
distortions, the time constant of the distance controller must be
shorter than the speed in the fast scanning direction / lateral
extent of the smallest feature to be resolved and necessarily
(much) larger than other time constants (RMS-to-DC, APIC, PLL). If
those conditions are satisfied, then its contribution to the
overall stability of the setup should be weak. This is what is
readily seen in figs.\ref{FIG_SCANPLL}, \ref{FIG_SCANAPIC} and, to
a certain extend in fig.\ref{FIG_SCANSPEED}, with reasonable
speeds. This is also confirmed by Couturier \emph{et al.} who have
recently found out that the distance controller had no effect on
the stability diagram in the self-excitation scheme and that only
the amplitude controller played a
role~\cite{note_virtual05,couturier05a}.

To summarize, this analysis has emphasized that a maximum of about
$15\%$ of apparent dissipation, mainly due to the PLL and not to
the APIC, could be generated (\emph{cf.} fig.\ref{FIG_SCANPLL}).
The contribution of the APIC, within the range of gains used, is
systematically smaller. We finally infer that, with our setup (UHV
and room temperature) and under typical experimental conditions
(scanning speed, PLL and APIC gains...) corresponding to
$E_{d_0}\simeq 2$~eV/cycle, apparent dissipation in the range of a
few eV \emph{per} cycle, as frequently reported in the literature,
is unlikely to occur. A striking example is provided by
R.~Hoffmann \emph{et al.}~[\onlinecite{hoffmann03a}] who put in
evidence a dissipation of 3~eV/cycle on a NiO(001) sample, despite
operating at 4~K~[\onlinecite{note_virtual10}]. This strongly
suggests the interpretation of the experimental damping images in
terms of ``physical" dissipation.

\section{Conclusion}
The realization and successful testing of a modular nc-AFM
simulator implemented with LabView$^{\text{TM}}$ is reported. The
design is based on a real electronics which includes a digital
PLL, the output of which is used to detect the frequency shift but
also to generate the time dependent phase of the excitation
signal. Good agreement is obtained between the locking behavior of
the real PLL and the PLL from the simulator. The optimum locking
time of the PLL is found to be about 0.35~ms. The behavior of the
amplitude controller is also found to correctly describe the real
setup with an optimum response of 2~ms. The analysis of the time
constants of the former two components provides evidence that the
electronics tracks properly the cantilever dynamics if the PLL
runs more than twice faster than the amplitude controller. When
the system is operated with properly chosen parameters, frequency
shift \emph{vs.} distance curves successfully compare to an
analytic expression which ignores the finite response of the
electronics. No phase deviation resulting in apparent dissipation
occur if the center frequency of the PLL tracks the resonance
frequency shifted by the tip-sample interaction. An estimate of
the minimum dissipation expected to be detected experimentally
gives some insights on the relevance of apparent dissipation which
can conditionally be generated numerically. This provides a
framework to discuss the overall contribution of apparent
dissipation during experiments. Computations of scan lines show
that when the system is operated with experimentally relevant
parameters, the contribution of the proportional and integral
gains of the amplitude controller and the scan speed (up to
$20$~nm.s$^{-1}$) do not lead to significant apparent dissipation.
To give orders of magnitude, the worse situation (frequency
tracker disengaged) leads to a maximum of $15\%$ more dissipation
than the intrinsic dissipation of the free cantilever. This is
below the values which are experimentally reported. This strongly
suggests that nc-AFM
damping images mainly reflect physical channels of dissipation and not electronics artifacts.

Besides Ref.\cite{couturier05a}, two publications dealing with the
implementation and/or performance of ``virtual force microscopes"
have appeared recently. These simulation codes are analogous to
ours but differ in detail. Kokavecz et al.\cite{kokavecz06a}
proposed and tested a numerical scheme designed to produce
response times of the whole simulated setup, as well as of
separate blocks (amplitude, phase and distance controllers) as
short as 0.1 ms. Trevethan et al.\cite{trevethan06a} used the
scheme described in Ref. \cite{polesel05} to compute
fingerprint-like responses in the frequency shift, the minimum
tip-sample distance and the damping signal caused by an
atomic-scale configuration change at the surface of the sample.
This change was first predicted from atomistic simulations and
then induced upon approach under distance control down to a
judiciously chosen frequency setpoint. A manual describing the
combined atomistic simulation\cite{kantorovich04a} and virtual
force microscope codes is now available
online\cite{note_virtual19}.

\acknowledgments The authors acknowledge the Swiss National Center
of Competence in Research on Nanoscale Science and the National
Science Foundation for financial support. They are grateful to
R.~Bennewitz (Mac Gill Univ., Canada), C.~Loppacher (Dresden
Univ., Germany), to colleagues of the electronics workshop of the
University of Basel for discussions and advices, in particular
Ch.~Wehrle H.-R.~Hidber and M.~Steinacher, to S.~Gauthier,
J.~Polesel-Maris and X.~Bouju (CEMES, France) for stimulating
discussions on noise, H.~Hug, N.~Pilet and T.~Ashworth (Basel
Univ.) for discussions on the frequency tracker and to
G.~Couturier for having provided a copy of his poster presented at
the 2004 nc-AFM conference (Seattle, USA).

\newpage
\bibliographystyle{unsrt}

\begin{thebibliography}{10}
\bibitem{note_virtual11}
{This can readily be seen by considering the simple case of the
free
  cantilever. If exciting at the resonance frequency, the oscillation amplitude
  matches the maximum of the Lorentzian curve, but if the excitation frequency
  changes, due to any reason, the amplitude decreases.}
\bibitem{note_virtual12}
{The images reported in the reference have been obtained out of a
force field
  calculated between an MgO tip and a CaF$_2$ slab.}
\bibitem{note_virtual16}
{This is an order of magnitude of the $\Delta f$ usually reported
in the
  litterature for cantilevers with $k=40$~N.m$^{-1}$ in UHV $(Q=30000)$ and at
  room temperature.}
\bibitem{note_virtual14}
{However, in the Van der Waals regime, the tip is relatively
distant from the
  surface. No atomic feature is expected to be observed. The bandwidth can
  therefore be made bigger accordingly.}
\bibitem{note_virtual07}
{On the base of the work by R. P\'erez with the Si(111)$5\times 5$
  (ref.[\onlinecite{perez98a}]), S.-H. Ke (Ch.16 in
  ref.[\onlinecite{morita02a}]), state that even at tip-surface distances of
  $\approx 5~\AA$, there is an onset of covalent chemical bonding between the
  dangling bonds on the tip and the surface.}
\bibitem{note_virtual10}
{According to the experimental parameters reported ($A_0=5$~nm,
  $k_c=43$~N.m$^{-1}$ and $Q=100000$), $E_{d_0}\simeq 0.2$~eV/cycle. Therefore
  the ratio between the maximum of dissipation to $E_{d_0}$ is $1500\%$! Even
  if $10\%$ of apparent dissipation is generated, this is still weak compared
  to the overall dissipation which is detected.}
\end{thebibliography}

\newpage
\section*{Appendices}
\subsection{The analytical method}\label{SECTION_ANALYTIC}
The analytical method gives analytical, tractable expressions of
the frequency shift versus distance upon the force expression.
Couple of years ago, F.J.~Giessibl and M.Guggisberg have proposed
approached expressions for the frequency induced by a Morse
potential~\cite{giessibl97a,guggisberg00a}. Here a complete
expression is provided. The following calculation is based on a
variational method extensively detailed in former
articles~\cite{nony01a}. We start from equation \ref{EQU_OTCS} and
postulate a solution of the differential equation where the
amplitude and the phase are not constant, but are allowed to vary
slowly within time, that is over durations much longer than the
oscillating period, namely~:

\begin{equation}
z(t)=A(t) \cos\left[ \omega t+ \varphi(t)\right]
\end{equation}

\noindent Equation \ref{EQU_OTCS} is then equivalent to~:

\begin{equation}\label{EQU_SYSREF}
\left\{
\begin{array}{ll}
\alpha \cos\left( \varphi(t)\right)-\beta \sin \left( \varphi(t)\right)=\gamma\\
\beta \cos\left( \varphi(t)\right)+\alpha \sin \left(
\varphi(t)\right)=0
\end{array}
\right.
\end{equation}

\noindent with~:

\begin{equation}
\left\{
\begin{array}{lll}
\alpha=\ddot{A}(t)+\frac{\omega_0}{Q}\dot{A}(t)+\left\{\omega_0^2-\left[ \omega+ \dot{\varphi}(t)\right]^2\right\}A(t)+\frac{2}{m}\tilde{F_c}\\
\beta=A(t)\ddot{\varphi}(t)+\left[2\dot{A}(t)+\frac{\omega_0}{Q}A(t)\right]\left[ \omega+ \dot{\varphi}(t)\right]+\frac{2}{m}\tilde{F_d}\\
\gamma=\omega_0^2K_3 A_{exc}
\end{array}
\right.
\end{equation}

\noindent $\tilde{F_c}$ and $\tilde{F_d}$ are the first Fourier
components of the conservative and dissipative forces experienced
by the tip over one oscillation period $T$, respectively~:

\begin{align}
& \tilde{F_c} = \frac{1}{T}\partial_A
\left\{\int_0^T{V_\text{int}(t)dt} \right\} \label{EQU_CONSCONTRIBUTION} \\
& \tilde{F_d} = \frac{1}{T A}\partial_\varphi
\left\{\int_0^T{V_d(t)dt}\right\} \label{EQU_DISSCONTRIBUTION}
\end{align}

\noindent Setting
$\dot{A}(t)=\ddot{A}(t)=\dot{\varphi}(t)=\ddot{\varphi}(t)=0$
yields to the steady equations of the oscillator in amplitude and
phase out of which the relevant variables, namely $\Delta f=f-f_0$
and driving amplitude, related to the damping, can be extracted~:

\begin{align}
& \tilde{F_c} = k_cA \left( \frac{\Delta f}{f_0}-\frac{f+2f_0Q\frac{\tilde{F_d}}{k_cA}}{2f_0Q\tan(\varphi)}\right) \label{EQU_CONSCONTRIBUTION2}\\
& \tilde{F_d} = \frac{k_c A
f}{2f_0}\left(\frac{K_3A_\text{exc}}{A\left( 1+\frac{\Delta
f}{f_0}\right) \left( 1+ 1/\tan(\varphi)\right)}-\frac{1}{Q}
\right) \label{EQU_DISSCONTRIBUTION2}
\end{align}

\noindent Equation \ref{EQU_CONSCONTRIBUTION2} illustrates that,
if the phase is properly maintained at $-\pi/2$~rad, the frequency
shift is essentially due to the conservative interaction, whatever
the dissipative contribution. Besides,
equ.\ref{EQU_CONSCONTRIBUTION2} gives the well known expression of
the frequency shift due to the conservative
contribution~\cite{giessibl00a}~:

\begin{equation}\label{}
\frac{\Delta f}{f_0}=\frac{\tilde{F_c}}{k_cA}
\end{equation}

\noindent The expression of $\tilde{F_c}$ related to the Morse
potential given in equ.\ref{EQU_FORCE} can be integrated
analytically~:

\begin{equation}\label{EQU_SINTMORSE}
S_\text{int}^\text{Morse}=\int_0^{\frac{2\pi}{\omega}}V_\text{int}^\text{Morse}(t)dt=-\frac{2\pi
U_0}{\omega}\left(2\Upsilon_{0, 1/\lambda}-\Upsilon_{0,
2/\lambda}\right),
\end{equation}

\noindent where~:

\begin{equation}
\Upsilon_{\alpha, \beta}=e^{-\beta(D-r_c)}\times
\text{BesselI}(\alpha,\beta A)
\end{equation}

\noindent $\text{BesselI}(\alpha,\beta A)$ is the modified Bessel
function of first kind for the parameters $\alpha$ and $\beta A$.
As shown by equ.\ref{EQU_CONSCONTRIBUTION}, the method not only
requires to estimate the action related to
$S_\text{int}^\text{Morse}$, but also its derivative with respect
to the amplitude $A$~:

\begin{equation}
\partial_A S_\text{int}^\text{Morse}=-\frac{4\pi U_0}{\omega\lambda}\left(\Upsilon_{1,
1/\lambda}-\Upsilon_{1, 2/\lambda}\right)
\end{equation}

\noindent Regarding the Van der Waals contribution, it was
demonstrated~\cite{nony01a}~:

\begin{equation}
\partial_A S_\text{int}^\text{VdW}=-\frac{\pi HRA}{3\omega \left(D^2-A^2\right)^{3/2}}
\end{equation}

\noindent The frequency shift due to both contributions can
finally be deduced~:

\begin{equation}\label{EQU_DELTAFANALYT}
\frac{\Delta f}{f_0}=-\frac{1}{k_cA}\left[ \frac{HRA}{6
\left(D^2-A^2\right)^{3/2}}+\frac{2
U_0}{\lambda}\left(\Upsilon_{1, 1/\lambda}-\Upsilon_{1,
2/\lambda}\right)\right]
\end{equation}

\subsection{APIC response time}\label{SECTION_APICTIMERESPONSE}
\begin{figure}[t]
 \includegraphics[height=3cm]{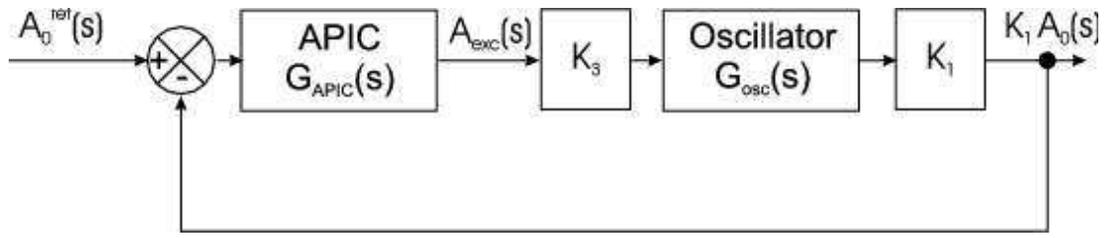}\\
 \caption{Simplified scheme of the closed loop for the characterization of the response time of the APIC.}\label{FIG_APICTHEO}
\end{figure}

The problem consists in analyzing the dynamics of the oscillator
the amplitude of which is controlled by a proportional-integral
controller. For that purpose, all the components are assumed to be
linear. No interaction occurs between the oscillator and the
surface. The contributions of the RMS-to-DC and of the band pass
filter are neglected. The system under investigation is given in
figure \ref{FIG_APICTHEO}.

\subsubsection{Transfer function of the closed loop}
We start from the set of equations \ref{EQU_SYSREF} wherein the
phase is assumed to be constant and fixed to
$\varphi(t)\rightarrow\varphi=-\pi/2$. This assumption is argued
by the PLL behavior which maintains the phase almost constant,
even when the interaction occurs (\emph{cf.} fig.
\ref{FIG_FORCEDISTANCE}(b)). The assumption implies~:
$\omega=\omega_0$, $\dot{\varphi}(t)=\ddot{\varphi}(t)=0$ and
obviously, we only focus at changes occurring in the resonance
amplitude $A(t)=A_0(t)$. The set of equations \ref{EQU_SYSREF} is
then equivalent to~:

\begin{equation}
\left\{
\begin{array}{ll}
\beta=\gamma\\
\alpha=0
\end{array}
\right.
\end{equation}

\noindent Keeping $\beta=\gamma$ yields to~:

\begin{equation}\label{EQU_DIFFSTEADY}
\dot{A}_0(t)=\frac{\omega_0}{2}\left\{ K_3
A_{exc}-\frac{A_0(t)}{Q}\right\}
\end{equation}

\noindent The transfer function of the block standing for the
oscillator can thus be written~:
\begin{equation}\label{EQU_TF_SYS}
G_{osc}(s)=\frac{K_1A_0(s)}{A_{exc}(s)}=K_1K_3\frac{b}{s+a}
\end{equation}

\noindent where~:

\begin{equation}
 \left\{
\begin{array}{ll}
b=\frac{\omega_0}{2}\\
a=\frac{\omega_0}{2Q}
\end{array}
\right.
\end{equation}

\noindent The transfer function of the APIC being $
G_{APIC}(s)=K_p^\text{ac}+K_i^\text{ac}/s$, the transfer function
of the closed loop $G_{cl}(s)=K_1A_0(s)/A_0^{ref}(s)$ can now be
calculated~:

\begin{equation}\label{EQU_TF_CLOSEDLOOP}
G_{cl}(s)=\frac{b\tilde{K}_p^\text{ac}s+b\tilde{K}_i^\text{ac}}{s^2+(a+b\tilde{K}_p^\text{ac})s+b\tilde{K}_i^\text{ac}},
\end{equation}

\noindent with~:

\begin{equation}\label{EQU_PIRESCALED}
 \left\{
\begin{array}{ll}
\tilde{K}_p^\text{ac}=K_1K_3K_p^\text{ac}\\
\tilde{K}_i^\text{ac}=K_1K_3K_i^\text{ac}
\end{array}
\right.
\end{equation}

\noindent The proportional and integral gains are scaled by the
transfer functions of the piezoelectric actuator and of the
photodiodes, $K_3$ and $K_1$, respectively.

\subsubsection{Analogy} Equation \ref{EQU_TF_CLOSEDLOOP} has two
poles~:

\begin{equation}
s_{1,2}=-c\pm\sqrt{c^2-b\tilde{K}_i^\text{ac}},
\end{equation}

\noindent where $c$ is the parameter given in equation
\ref{EQU_C}, which can also be written~:

\begin{equation}\label{EQU_C2}
 c=\frac{a+b\tilde{K}_p^\text{ac}}{2}
\end{equation}

\noindent Equation \ref{EQU_TF_CLOSEDLOOP} is thus almost analog
to a standard $2^{nd}$ order system, the canonical form of which
can be written~:

\begin{equation}
G(s)=\frac{\omega_n^2}{s^2+2\zeta\omega_n s+\omega_n^2},
\end{equation}

\noindent where $\omega_n$ and $\zeta$ are the characteristic
frequency and damping factor of the system, respectively. Thus~:

\begin{equation}
\zeta=\frac{c}{\sqrt{b\tilde{K}_i^\text{ac}}}
\end{equation}

\noindent Now, it's well known that the position of $\zeta$ with
respect to 1 defines the overall behavior of the system~:

\begin{equation}
\left\{
\begin{array}{lll}
\text{Undercritically damped regime }\Rightarrow \zeta<1
\Leftrightarrow c<\sqrt{b\tilde{K}_i^\text{ac}}\\
\text{Critically damped regime }\Rightarrow \zeta=1
\Leftrightarrow c=\sqrt{b\tilde{K}_i^\text{ac}}\\
\text{Overcritically damped regime }\Rightarrow \zeta>1 \Leftrightarrow c>\sqrt{b\tilde{K}_i^\text{ac}}
\end{array}
\right.
\end{equation}

\subsubsection{Analysis to a step response} To assess how fast the
controller reacts, we investigate the response of the controller
to a step in amplitude upon it is in the over-, under- or
critically damped regime. Let's assume a step of amplitude $A_s$,
the corresponding transfer function is $G_s(s)=A_s/s$ and the
transfer function of the closed loop system to which the step is
applied is therefore $G_{cls}(s)=G_s(s)\times G_{cl}(s)$, that
is~:

\begin{equation}
G_{cls}(s)=\frac{A_s b \tilde{K}_p^\text{ac}}{\left(s+c
\right)^2-c^2+b\tilde{K}_i^\text{ac}}+\frac{A_s b
\tilde{K}_i^\text{ac}}{\left(s+c
\right)^2-c^2+b\tilde{K}_i^\text{ac}}\times \frac{1}{s}
\end{equation}

\noindent The time-dependent solutions
$g_{cls}(t)=\mathcal{L}^{-1}\{G_{cls}(s)\}$ are~:

\begin{itemize}
    \item Overcritically damped regime~: $c>\sqrt{b\tilde{K}_i^\text{ac}}$\\
    \begin{equation}\label{EQU_GCLS_OVER}
    g_{cls}(t)=A_s\left\{1+\varsigma_{-}e^{-(c+\xi)t}
    -\varsigma_{+}e^{-(c-\xi)t} \right\}
    \end{equation}
    \noindent with $\varsigma_{\pm}= (c\pm \xi-b\tilde{K}_p^\text{ac})/2\xi$ and $\xi=\sqrt{c^2-b\tilde{K}_i^\text{ac}}$.

    \item Critically damped regime~: $c=\sqrt{b\tilde{K}_i^\text{ac}}$
    \begin{equation}\label{}
    g_{cls}(t)=A_s\left\{
    1-e^{-ct}+\left[ b\tilde{K}_p^\text{ac}-c\right]te^{-ct}
    \right\}
    \end{equation}

   \item Undercritically damped regime~: $c<\sqrt{b\tilde{K}_i^\text{ac}}$
   \begin{equation}\label{}
   g_{cls}(t)=A_s\left\{ 1-e^{-ct}\times \left[ \cos\left(\xi' t
   \right)+\frac{c-b\tilde{K}_p^\text{ac}}{\xi'}\sin \left( \xi'
   t\right)\right]\right\}
   \end{equation}
   \noindent with $\xi'=\sqrt{b\tilde{K}_i^\text{ac}-c^2}$.
\end{itemize}

\subsubsection{Summary} The transition from the over to the under
critically damped regime is thus controlled by the parameter $c$,
that is the proportional gain, and occurs when the condition
$c=\sqrt{b\tilde{K}_i^\text{ac}}$ is fulfilled. In the critically
damped regime, the relationship between the two gains is~:

\begin{equation}\label{EQU_TRANSITION}
\tilde{K}_i^\text{ac}=\frac{\omega_0}{8}\left(\frac{1}{Q}+\tilde{K}_p^\text{ac}\right)^2
\end{equation}

\noindent The time constant of the system can be extracted upon
the regime and the time scale that are considered. For short time
scales, the response time of the controller is typically $1/c$ in
the critically damped regime and
$t_\text{resp}=1/(c+\sqrt{c^2-b\tilde{K}_i^{ac}})$ in the
overcritically damped regime (\emph{cf.} equ.\ref{EQU_GCLS_OVER}),
which is given in equation \ref{EQU_TIMERESPONSEAPIC}. Figure
\ref{FIG_APICTIMERESPONSE} shows that the response times of the
real system and of the simulated machine reasonably match
$t_\text{resp}$ while the RMS-to-DC converter has a negligible
influence on the system dynamics.

\newpage
\section*{Figures}

\end{document}